\documentclass[geosciences,review,accept,oneauthor,pdftex,10pt,a4paper,]{mdpi} 
\usepackage{wrapfig}
\usepackage{booktabs} 
\usepackage{multirow}
\usepackage{soul} 
\usepackage{microtype}
\usepackage{upgreek}
\newcommand{\angstrom}{\mbox{\normalfont\AA}}
\firstpage{1} 
\articlenumber{0}
\pubvolume{00}
\pubyear{2019}
\copyrightyear{2019}
\history{Received: 17 December 2018; Accepted: 17 January 2019; Published: 4 March 2019}
\issuenum{0} 
\updates{yes}

\Title{The Impact of Stellar Surface Magnetoconvection and Oscillations on the Detection of Temperate, Earth-Mass Planets Around Sun-Like Stars}

\Author{H~M. Cegla $^{1,\dagger}$\orcidA{}}
\AuthorNames{Heather~M. Cegla}

\address{%
$^{1}$ \quad Observatoire de Gen\`eve, Universit\'e de Gen\`eve, 51 Chemin des Maillettes, 1290 Versoix, Switzerland; h.cegla@unige.ch\\
$^{\dagger}$ \quad CHEOPS Fellow}

\abstract{Detecting and confirming terrestrial planets is incredibly difficult due to their tiny size and mass relative to Sun-like host stars.~However, recent instrumental advancements are making the detection of Earth-like exoplanets technologically feasible. For example, Kepler and TESS photometric precision means we can identify Earth-sized candidates (and PLATO in the future will add many long-period candidates to the list), while spectrographs such as ESPRESSO and EXPRES (with~an~aimed radial velocity precision [RV] near 10~cm~s$^{-1}$) mean we will soon reach the instrumental precision required to confirm Earth-mass planets in the habitable zones of Sun-like stars. However, many~astrophysical phenomena on the surfaces of these host stars can imprint signatures on the stellar absorption lines used to detect the Doppler wobble induced by planetary companions. The result is stellar-induced spurious RV shifts that can mask or mimic planet signals. This review provides a brief overview of how stellar surface magnetoconvection and oscillations can impact low-mass planet confirmation and the best-tested strategies to overcome this astrophysical noise. These noise reduction strategies originate from a combination of empirical motivation and a~theoretical understanding of the underlying physics. The most recent predications indicate that stellar oscillations for Sun-like stars may be averaged out with tailored exposure times, while granulation may need to be disentangled by inspecting its imprint on the stellar line profile shapes. Overall, the literature suggests that Earth-analog detection should be possible, with the correct observing strategy and sufficient data collection.
}


\begin{document}

%
\section{Introduction}
The confirmation and detailed characterization of exoplanets necessitates a measurement of the planetary mass. At present, the most tried and tested technique to obtain such a mass measurement comes from analyzing the planetary-induced Doppler wobble of the host star (arising because the pair share a common center of mass). For a true Earth-analog, the planet-induced radial velocity (RV) on the host star is a minuscule $\sim$9~cm~s$^{-1}$---making this feat extremely challenging; hence, why~such a confirmation still eludes us to date. However, recent advancements in instrumentation promise to finally make this challenge technologically feasible. For example, both the ESPRESSO \citep{pepe14, gonzalez17}  and EXPRES \citep{jurgenson16, fischer17} spectrographs have recently had first light, with ESO open time on ESPRESSO commencing autumn 2018 (ESPRESSO: Echelle SPectrograph for Rocky Exoplanets and Stable Spectroscopic Observations; EXPRES: EXtreme PREcision Spectrometer). ESPRESSO is aiming to achieve $\sim$10~cm~s$^{-1}$ precision for (bright) targets in the Southern Hemisphere, while EXPRES hopes to approach this regime for similar targets in the North. However, trying to determine the RV of the host star can often be difficult due to inhomogeneities on the stellar surfaces themselves. For stabilized spectrographs, these RVs are determined by observing the stellar absorption lines in a relatively large passband, usually several thousand angstroms wide, and cross-correlating them with a template mask. The presence of a planetary companion induces wholesale Doppler shifts of the individual lines, where the net influence can be determined from the center of the cross-correlation function (CCF) \cite{figueira18}. However, if the stellar lines change shape this will also be reflected in the CCF. For instance, a dark starspot (or bright plage/facular region) creates an emission bump (absorption dip) in the stellar lines, and this asymmetry alters the center-of-light for both the individual line profiles and the overall CCF; as a result, spurious RV shifts are measured and can be confused for or mask the induced shifts from a planetary companion \citep{saar97}. For a magnetically active star, the spots or plage/faculae can induce RV shifts on the 1--100 m~s$^{-1}$ level \citep{saar97}; hence, the very active stars are often avoided when trying to search for Earth-analog companions. Nonetheless, even~very magnetically quiet stars will still suffer anomalies from inhomogeneities arising from both stellar surface magnetoconvection and oscillations on the 0.1--1 m~s$^{-1}$ level \citep{schrijver00}. Consequently, it is essential that we also understand these low-amplitude signals, as they are still large enough to completely swamp the signal from an Earth-twin and other low-mass, long-period planetary companions. There~are numerous works dedicated to disentangling and removing the larger amplitude astrophysical signals from high precision RVs (e.g., see \cite{saar97, hatzes02, bonfils07, desort07, melo07, queloz09, dumusque11b, aigrain12, boisse09, boisse11, dumusque14b, hatzes10, haywood14, haywood16, hebrard16, herrero16, lopezmorales16, meunier10a, meunier10b, meunier17}, and references therein). The focus of this paper is to briefly discuss the physics driving convection and oscillation-induced RV variability (Section~\ref{sec:conv_osc}), review the recent works in the literature that aim to reduce these stellar noise sources in high precision RVs (Section~\ref{sec:strat}), and comment on the prospects for the future confirmation of habitable worlds (Section~\ref{sec:future}).

\section{Stellar Surface Magnetoconvection and Oscillations}
\label{sec:conv_osc}
\begin{wrapfigure}[22]{r}{0.45\textwidth}
\vspace{-24pt}
\centering
\includegraphics[trim=0.2cm 0 0.3cm .5cm, clip, scale=0.4]{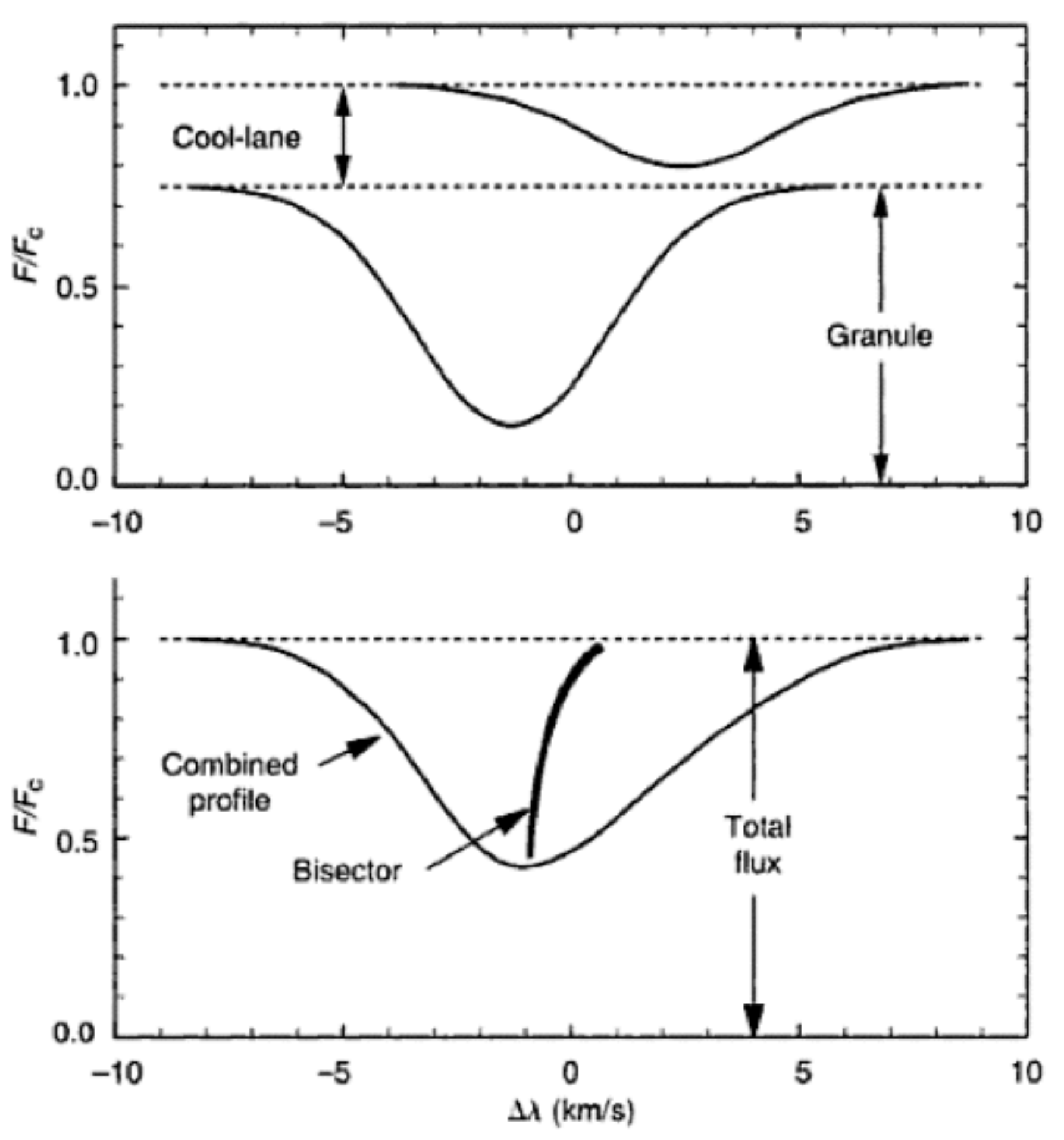}
\vspace{-20pt}
\caption{Schematic from \cite{gray05} displaying the relative (top) and combined (bottom) contribution of the bright, blueshifted granules and the dark, redshifted intergranular lanes; note the lane profile has been artificially offset in flux for display purposes. The curvature in the line bisector displays the asymmetries in the combined profile.}
\vspace{-10pt}
\label{fig:gray_bi}
\end{wrapfigure}
Convection creates time-variable asymmetries in the stellar absorption lines due to a combination of upward flowing, bright, hot, blueshifted bubbles of plasma ($\sim$1~Mm in diameter) known as granules that eventually cool, darken, and fall back down, redshifting, into the surrounding regions known as intergranular lanes; since the granules are brighter and cover more surface area, they do not completely cancel out the intergranular lane contribution, which acts to depress the redward wing of the combined line profile---see Figure~\ref{fig:gray_bi}. Moreover, the larger contribution from the granules means the center-of-light for most absorption lines in a Sun-like star have an overall net blueshift (for the Sun this value is near 350~m~s$^{-1}$ \cite{gray05}). Magnetic fields can inhibit the convection, and over a magnetic activity cycle the net convective blueshift varies on the $\sim$10~m~s$^{-1}$ level \citep{meunier10a, dumusque14b,haywood16}. See \cite{meunier17a,meunier17b} for how this varies with spectral type, age, and magnetic activity, and see \cite{meunier13, meunier17, haywood16} for how we may correct for this type of large-amplitude stellar noise; this review will focus on the shorter timescale variability.

As individual granules evolve, with lifetimes of $\sim$5--6 min on the Sun, the ratio of granules to intergranular lane is constantly changing, which means the stellar line asymmetries, and therefore net RV shift, are also constantly changing. Typical velocities for individual plasma flows are 1--4~km~s$^{-1}$, but over the stellar disc much of the upflows and downflows cancel out, leaving net variability on the 0.1--1~m~s$^{-1}$ level for Sun-like stars \citep{schrijver00}.

\begin{figure}[t]
\centering
\includegraphics[scale=0.3]{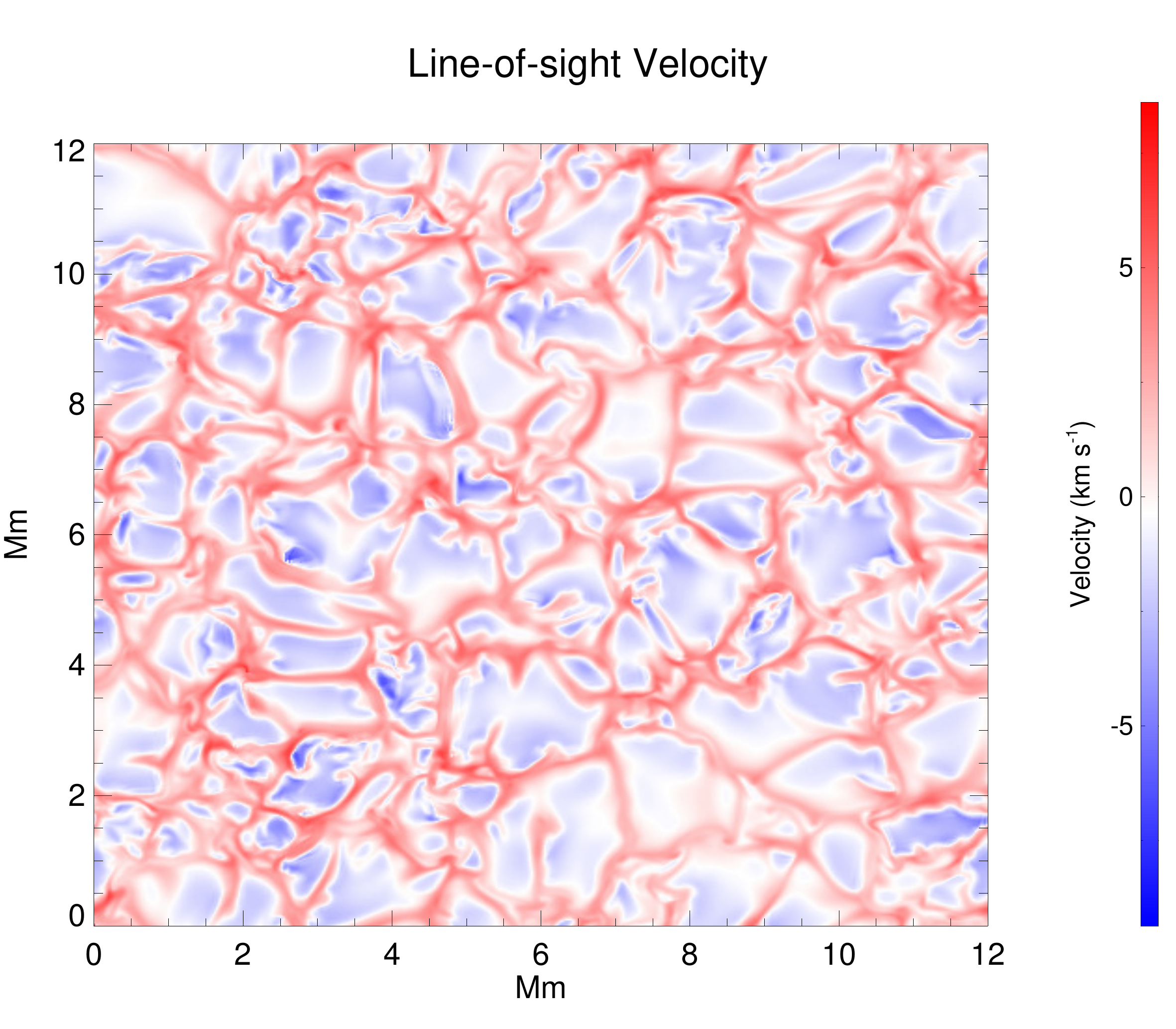}
\includegraphics[scale=0.3]{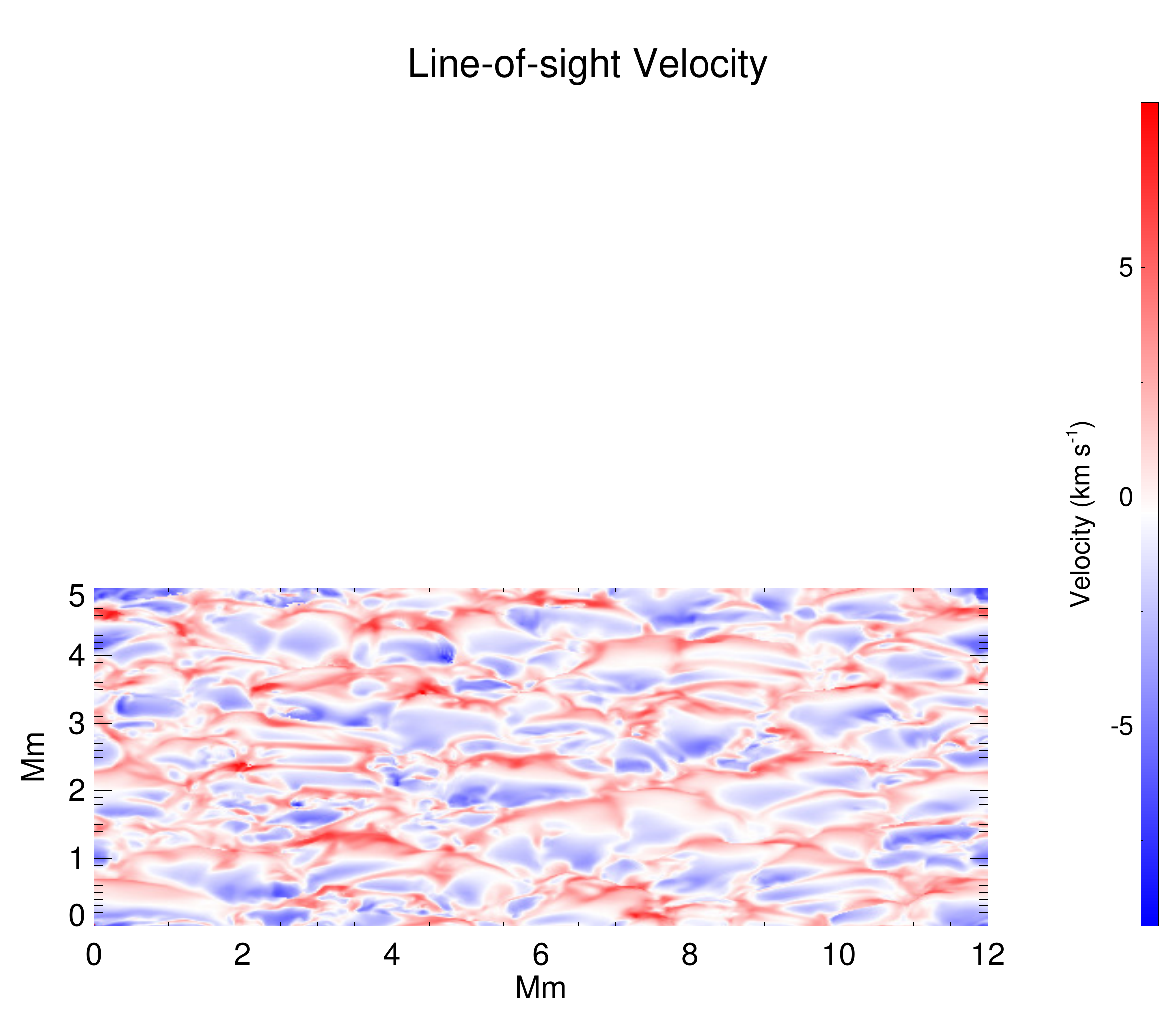}
\caption{Line-of-sight RVs for a 3D MHD simulation snapshot at disc center (left) and $\upmu$ = 0.5 (i.e., 60$^{\circ}$; subset of Figure~1 in \cite{cegla18a}). Negative and positive velocities denote blueshifts and redshifts, respectively.} 
\label{fig:2_80_vel}
\end{figure}
The presence of granulation on a spherical host star makes the surface corrugated; an apt analogy is to think of the granules as `hills' and the intergranular lanes as `valleys' \cite{dravins08}---standing at the top of a hill looking out to other hills provides a vastly different vantage point than a birds-eye view from above.~Near the limb it is impossible to see to the very bottoms of the intergranular lanes, granular walls become visible and only the near edge of their tops can be seen, and granules in the forefront can obstruct components in the background etc. Moreover, the plasma flows in a variety of directions, and flows that were once orthogonal at disc center have a line-of-sight component near the limb---see Figure~\ref{fig:2_80_vel}. This means the average line profile shapes and positions also change as a~function of center-to-limb. In fact, near the limb the net convective blueshift seen at disc center can disappear completely, and can even redshift \citep{dravins82}. Part of this effect is due to a smaller projected area near the limb, but the main driver is that velocity flows moving away from the observer are more often seen in front of the hotter plasma above the intergranular lanes and those moving towards the observer become increasing blocked by granules in the forefront \citep{balthasar85, asplund00}. The amplitude and exact nature of these effects will depend on the line properties and magnetic field strength; e.g., shallower lines, experiencing stronger magnetic fields, will have some of the least center-to-limb variations \citep{cegla18a}. If~this limb-dependent convective blueshift is ignored, it~may significantly impact transit observations, such~as the Rossiter-McLaughlin effect \citep{shporer11,cegla16a} and potentially transmission spectroscopy \citep{apai18}; hence, it~could be important for planet characterization, e.g., to further assess habitability. 

Individual granules cluster together in large groups or cells known as supergranules, with~diameters around $\sim$35~Mm. Plasma flows radially away from the centers of these cells with velocities of a few hundred m~s$^{-1}$, much slower than the km~s$^{-1}$ rates of the smaller, individual granular cells. The origin for this phenomenon remains a puzzle, {and has only been studied in detail on the Sun,} but it is thought to be buoyantly driven and may be related to thermal convection---see reviews by \cite{rieutord10,rincon18} for more details. Observations and numerical simulations show that families of individual granules splitting can diffuse and advect magnetic flux towards the boundaries of the supergranular cells; as these families of granules interact, they can generate horizontal flows, which may contribute to organization of the supergranules (\citep{roudier16, malherbe18, rieutord10, rincon18}, and references therein). The slower flow velocities, combined with projection effects, means the overall RV variability from supergranulation is lower in amplitude that from the small-scale granulation. However, the supergranule cells have much longer lifetimes, around 1.6--1.8 days for the Sun. Even though supergranulation may be lower in amplitude than small-scale granulation, its longer lifetime means it may still pose issues when searching for low-mass, long-period planets. 

\begin{figure}[t]
\centering
\includegraphics[trim = .3cm 0 0 0, clip, scale=0.41]{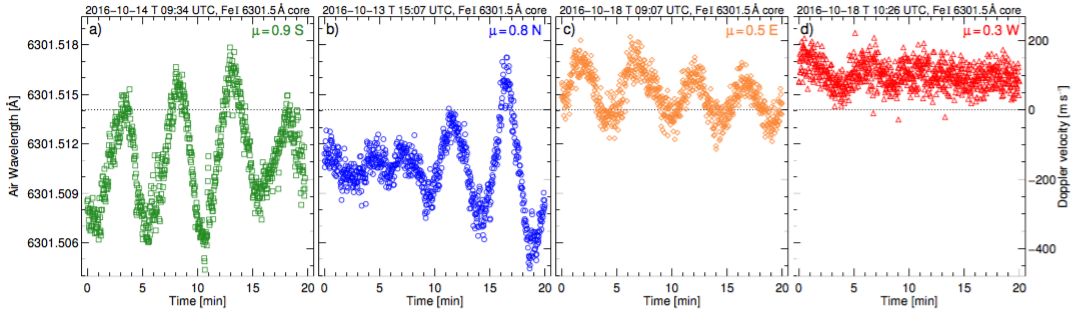}
\caption{Center-to-limb solar observations of the Fe~I~6302~$\angstrom$ line taken with the LARS spectrograph (Figure~4 from \cite{lohner18}), covering a region 10'' wide; the scale on the right-hand side shows how the oscillations change in amplitude from near 200~m~s$^{-1}$ at disc center, down to $\sim$50~m~s$^{-1}$ near the limb.}
\label{fig:LB_osc}
\end{figure}
\begin{wrapfigure}[14]{r}{0.55\textwidth}
\centering
\vspace{-10pt}
\includegraphics[trim = .2cm 0.4cm 0 0.9cm, clip, scale=0.5]{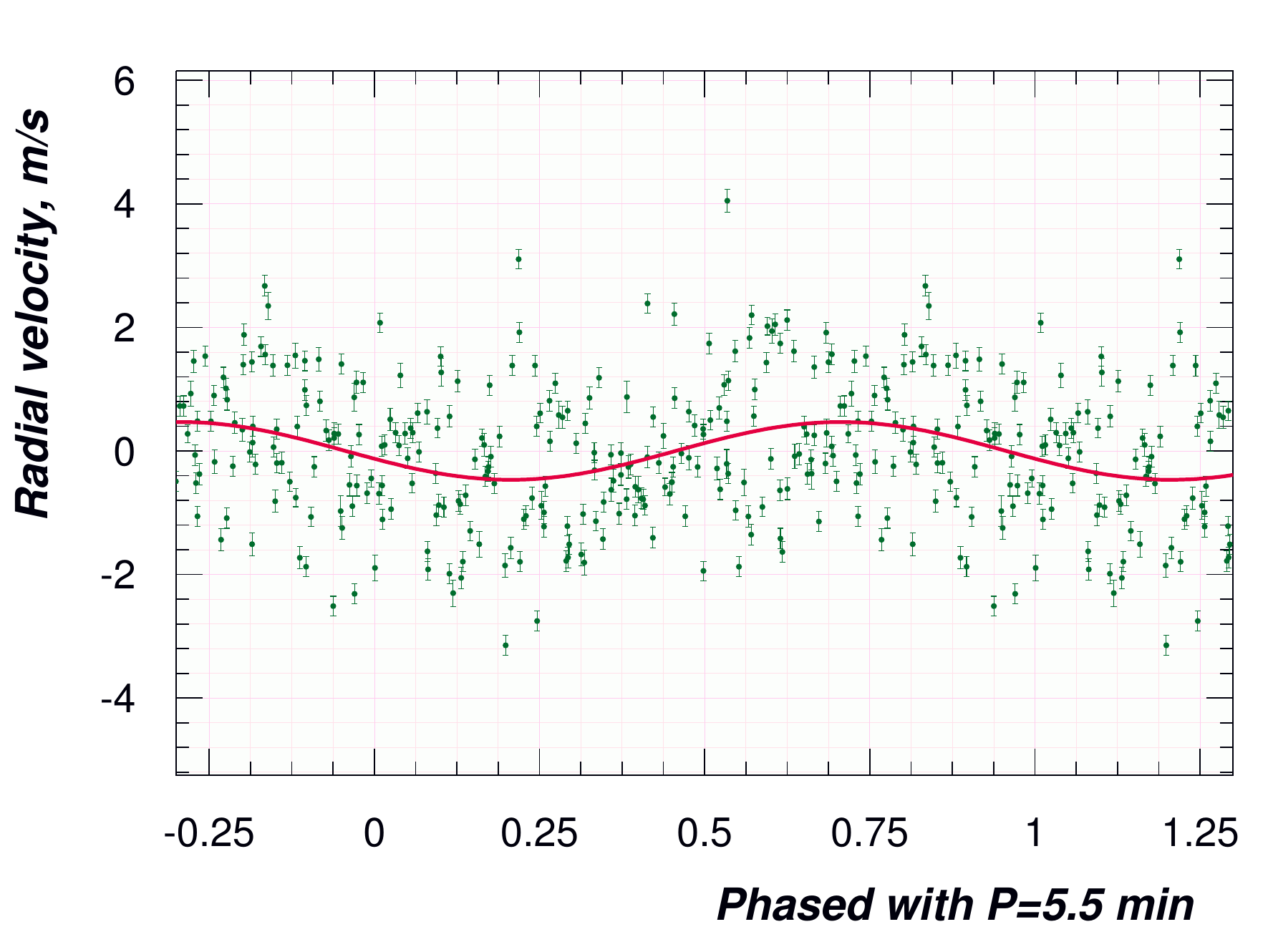}
\vspace{-20pt}
\caption{Sun-as-a-star observations from the PEPSI spectrograph, phased on the peak solar p-mode oscillations (subset of Figure~7 from \cite{strassmeier18}).}
\label{fig:pepsi_osc}
\end{wrapfigure}

The motions of the plasma flows in a given stellar convective envelope excite acoustic waves that constructively and destructively interfere with one another to create stochastic standing waves, with~lifetimes near $\sim$5 min for Sun-like stars. The most significant restoring force in the near-surface layers arises from pressure gradients and the resulting resonant modes are aptly termed p-modes (with~the `p' standing for pressure \cite{chaplin13}). The influence of the p-modes on the observed line profiles depends on the position across the stellar disc; observations near the limb see physically higher layers in the solar atmospheres, as well as contributions from velocity flows that are both vertical and horizontal (with~respect to disc center). Solar observations shown in Figure~\ref{fig:LB_osc} display how the RV amplitudes from a small patch on a stellar disc can change from 100--200~m~s$^{-1}$ at the center to $\sim$50~m~s$^{-1}$ or less near the limb{, and how the oscillation-induced RV variability dominates over the granulation in local patches}. Over the whole stellar disc the various oscillations average down to the m~s$^{-1}$ level for Sun-like stars; this can be seen in the recent solar observations in Figure~\ref{fig:pepsi_osc}, where the well-known `5-min' p-modes can be seen with an amplitude near 0.5~m~s$^{-1}$ and a root-mean-square (rms) just over 1~m~s$^{-1}$. Part of the scatter in Figure~\ref{fig:pepsi_osc} is due to the stochastic nature of the oscillations (i.e. modes constructively and destructively interfering with one another), but part of it is also due to the contribution from granulation. The exact details of how (super-)granulation amplitudes and timescales compare with oscillation amplitudes/timescales across the HR diagram remains an open question. However, it is generally agreed that the amplitude of the oscillations in Doppler velocities tend to be higher; moreover, the variability from both (super-)granulation and oscillations increases as stars evolve since the granules become larger, with longer turnover timescales. Nonetheless, it is clear that signals from a combination of convection and oscillations can completely swamp the 9~cm~s$^{-1}$ amplitude from an Earth-analog. Moreover, even though these stellar phenomena occur on very different timescales to the orbit of an Earth-analog, their frequency structure will not be finely sampled by typical RV observations of exoplanets; hence, it would likely be difficult to try to separate them in the power spectrum or even to include them as a correlated noise term when modelling the RVs \citep{haywood16}. Hence, it is important to try to remove and/or disentangle these sources of stellar variability.

\section{Noise Reduction Strategies}
\label{sec:strat}
\textls[-15]{Stellar surface convection and oscillations have been studied intensely for many years; see~\cite{stein12, nordlund09, rieutord10, gizon05, basu16}} for comprehensive reviews. However, it was not until the dawn of the exoplanet era in the 1990s and the realm of high precision RVs, nearly 20 years later, that these very interesting signals have been viewed as a nuisance that must be removed from the data. {Historically, empirical (exoplanet-focused) studies have typically rolled these signals into a single `jitter' term, that often also include variations induced by starspots and plage/faclue, as well as even instrumental noise (e.g.,~see~\citep{santos00, saar03, wright05, cegla14a, bastien14, marchwinski15, yu18}). These~works may be useful to determine the optimal candidates for RV follow-up, but they do not explicitly focus on removing or disentangling these stellar sources from the Doppler-reflex motion induced by planetary companions. Additionally}, as the astrophysical signals {from oscillations and granulation} are much lower in amplitude than those from the magnetically active regions, there have been relatively few attempts to remove their signatures. In particular, the current generation of instruments (HARPS [High Accuracy Radial Velocity Planet Searcher], HARPS-N, HIRES [High Resolution Echelle Spectrometer] etc.) have an instrumental precision around 50~cm~s$^{-1}$, which is near the boundary of detection for convection and oscillations; this means most stellar `noise' removal strategies have only started to scratch the surface of these lower amplitude effects. Nonetheless, it is clear that the next generation of spectrographs will demand a proper treatment of astrophysical phenomena down to the cm~s$^{-1}$ level to reach their full potential. 

\subsection{Empirically Driven Strategies}
The first strategies to tackle convection and oscillations as a noise source in exoplanet RV data were empirically driven. Foremost, and perhaps most well-known and used to date, Ref.~\cite{dumusque11a} worked to optimize the observational strategy to simply average out or `beat down' the convection and oscillation noise. To determine the optimal observing strategy, Ref.~\cite{dumusque11a} created magnetically quiet model stars (i.e., with only convection and oscillations as noise sources, no spots, or plage/facular regions) based on an asteroseismic data set from the HARPS spectrograph. The asteroseismology only span 5--8 days, but this should be sufficient to characterize the convection and oscillations as their timescales are on the order of minutes for their target sample of G and K dwarfs (up to 1--2 days if also considering supergranulation effects). The observed RVs were transformed in Fourier space and the velocity power spectrum density (VPSD) was calculated for each target. Then the authors performed a fit to the p-mode oscillations and granulation (they considered both granulation and supergranulation, as well as the debated mesogranulation; note, there is a long-running debate in the literature about whether or not mesogranular flows exist as a distinct scale of convection -- see~\citep{nordlund09, rieutord10}, and references therein). The granulation model fits were governed by empirical solar laws derived by \cite{harvey84,andersen94,palle95}, that correspond to an exponentially decaying function. For the p-modes, Ref.~\cite{dumusque11a} was only interested in the single hump of excess power in the VPSD and not the individual modes; this~hump is well described by a Gaussian with full-width half-maximum equal to four times the large separation of the p-modes. Convolving such a Gaussian with the VPSD then allowed \cite{dumusque11a} to fit the p-modes with a Lorentzian function following \cite{kjeldsen05, arentoft08}. Finally, randomizing the phase of each frequency in the inverse Fourier transform, and then returning to RV space meant the authors could calculate/predict oscillation and granulation-induced RVs (with timescales of a few days) for any given period in time. {It is important to keep in mind, the granulation components are constructed to fit the background in the VPSD, and it is difficult to ascribe a physical meaning to them. For example, the mesogranulation component may be attributed to an artefact introduced by averaging procedures and is now largely considered to not be a distinct convection scale \citep{rieutord10, rincon18}. Moreover, supergranulation has only been concretely confirmed on the Sun and very little is known about its manifestation on other stars. In~fact, it is possible that this background component could also be due to the occurrence of magnetic bright points, the presence of faculae, or from variations in the small-scale granulation properties (see~\citep{kallinger14}, and references therein). On top of this, the combination of finite mode lifetimes and variations due to magnetic field fluctuations means there is an inherent variability that is not captured by the empirical dataset used by \cite{dumusque11a}; based~on Kepler observations, this may mean an extra $\sim$15\% uncertainty or more~\citep{kallinger14}. Nonetheless, this~approach still provides a reasonable starting point for exploring granulation and oscillations on Sun-like stars.} 

\begin{figure}[t]
\centering
\includegraphics[scale=0.64]{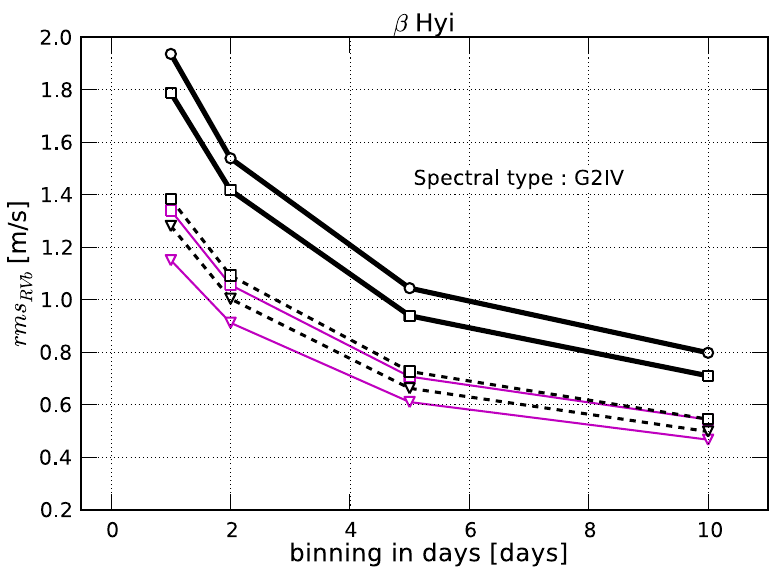}
\includegraphics[scale=0.64]{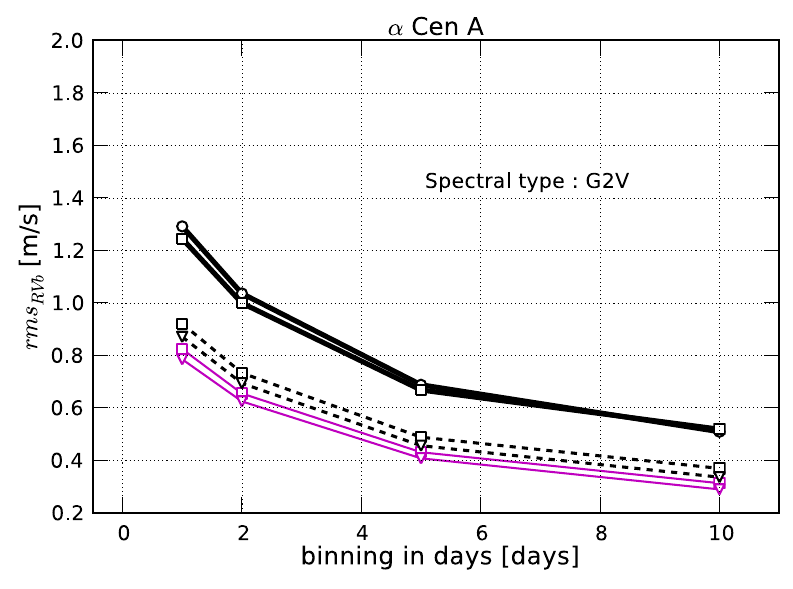}
\includegraphics[scale=0.645]{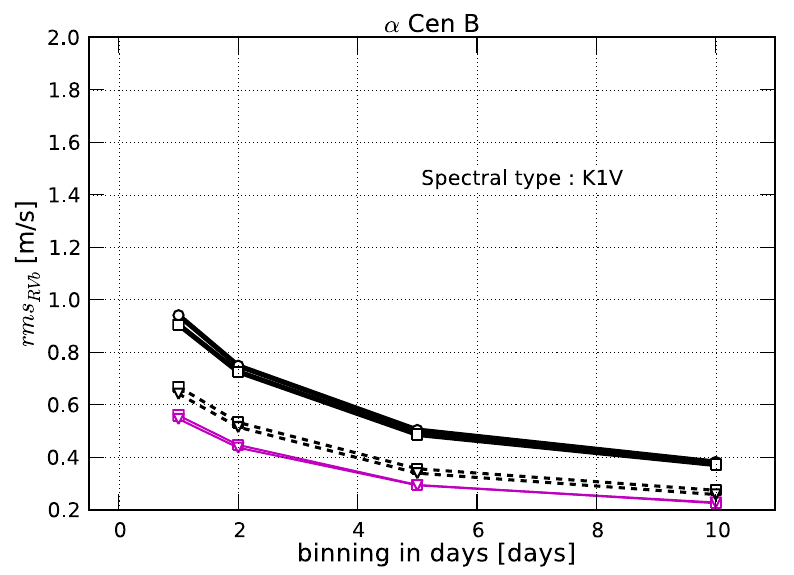}
\caption{A subset of Figure 4 from \cite{dumusque11a}, showing the {predicted} reduction in RV rms for various observational strategies {based on a simulated} subgiant and dwarf G star and a K dwarf. The thick, black curve represents one measurement per night, the dashed curve represents two per night, and the thin, purple curve represents three per night. Circles represent a total {(simulated)} observing time per night of 15 min. (1 exposure), squares are a total of 30 min per night (1, 2 or 3 equal length exposures), and the triangles 60 min per night (either 2 or 3 equal length exposures).}
\label{fig:dum_avg}
\end{figure}

Armed with these empirical fits, Ref.~\cite{dumusque11a} created artificially noisy, yet magnetically quiet, model~stars representative for G and K stars; the authors assumed an instrumental error equivalent to HARPS and that 256 nights would be available over the course of four years, on par with expectations for an intensive survey with a HARPS-like instrument (keeping in mind that most stars are only visible part of the year and that $\sim$20\% of the time can be expected to be lost due to bad weather). To begin, the authors tested the previous HARPS-GTO (Guaranteed Time Observations) strategy of 15 min exposures on $\sim$10 consecutive nights per month over the 4 years. Next, the authors simply tried doubling and quadrupling the total exposure time, and/or doubling and tripling the total number of observations in a given night; in other words they tried: 1 observation per night with a 15 min exposure (previous standard), 1 observation per night of 30 min, 2 observations per night of 15 min, 2~observations per night of 30 min, 3 observations per night of 10 min, and 3 observations per night of 20 min; in addition each strategy was tested for further binning the data together from 1--10 consecutive nights. See Figure~\ref{fig:dum_avg} for an example of the results for a subgiant and dwarf G star, as well as a K dwarf.   

Ref.~\cite{dumusque11a} found that doubling the exposure time had a very small effect on the overall noise reduction, while taking multiple observations in a given night did help to reduce the stellar noise more significantly. In particular, the authors claim they can remove the p-mode signatures with 15~min exposures and the granulation signature with a total of exposure time of 30 min---at least to the $\sim$50~cm~s$^{-1}$ level of their instrumental precision.~For meso- and supergranulation, they argue separating the observations in a given night by 2 and 5 h, respectively, can help to average out these noise sources. Overall, Ref.~\cite{dumusque11a} concluded that the best strategy for most Sun-like stars was 3~observations a night with 10 min exposure times, separated by 2 h for 10 consecutive days each month that the star is visible---as shown in Figure~\ref{fig:dum_avg}. One advantage of this strategy is the ease to universally apply it to all similar stars; however, it is important to keep in mind that once we can routinely reach a 10~cm~s$^{-1}$ instrumental precision this strategy may need to be adapted. Nonetheless, Ref.~\cite{dumusque11a} predicted they could reach an overall RV precision of $\sim$30~cm~s$^{-1}$ for a Sun-like star and possibly down to $\sim$20~cm~s$^{-1}$ for a~K dwarf with just 10 days of binning. {It is important to note that these predicted precisions are only applicable to the brightest targets, and the requirement for such intense monitoring would shrink the available target list even further.} 

Based on a bootstraping technique, Ref.~\cite{dumusque11a} calculated the false alarm probability (FAP) for a~variety of detection limits \citep{endl01, efron98}. They predicted that increasing the number of observations per night to 2 or 3 should push their detection limits down to planets $\sim$2--3 times as massive as the Earth. In~particular, since the K dwarfs have naturally lower amplitudes in convection and oscillations, they~argue that for these targets they may be able to detect habitable zones planets with a mass twice that of the Earths. However, it is not yet clear what is the true habitability and/or structure of planets in this mass regime. Moreover, it is apparent that the elusive Earth-analog would remain beyond the realm of realistic possibilities, at least in the 4 years of observational time considered here. 

More recently, Ref.~\cite{meunier15} tested similar noise reduction strategies, but with a different underlying model setup, {based off the Sun}---focusing specifically on {solar} granulation and supergranulation effects. In this case, the model star is constructed by populating a stellar disc with uprising granules and down-falling intergranular lanes, each with individual velocities. The granule size distributions, lifetimes, birth/death, splitting, and merging were all governed by empirical laws derived from the Sun \citep{roudier86,hirzberger97,hirzberger99}. The velocities of each granule/intergranular lane were based on distributions in the hydrodynamical (HD) solar simulation from \cite{rieutord02}, which spanned roughly an hour of physical time, with a $\sim$20 s cadence and had a physical box size of 30~$\times$~30 Mm$^2$ in the horizontal direction (and~3.2~Mm in the vertical direction); note that the variations from the p-modes were filtered from the HD simulation so that the noisy model stars only contain RV variations from granulation phenomena. It is also important to note that the HD simulation was constructed only at disc center, but the authors did use the distributions of the vertical and horizontal velocity flows to extrapolate the impact of projection effects across the stellar disc; however, they are likely missing some effects from the corrugated nature of granulation (e.g., granules will obstruct other granules near the limb and some velocities flows will be visible underneath the smaller granules etc.). {Additionally, as this work is based on the Sun, the laws describing the granulation properties may be difficult to extrapolate to other spectral types, where the plasma properties differ (e.g., see the simulations in \citep{beeck13a,beeck13, beeck15a, beeck15b}, wherein hotter stars show longer lived, faster flowing granules, with greater granular-intergranular lanes contrasts etc.).} Nonetheless, an advantage of this approach is that it should not be hindered by the current level of instrumental precision in a typical exoplanet-hunting spectrograph. 

Unfortunately, despite their large horizontal size the HD simulations did not include supergranulation; this might have been due to the shallow simulation box size, the lack of magnetic fields, or the time-series may have been too short for the supergranules to form \cite{rieutord02}.~As such, Ref.~\cite{meunier15} used empirical relations from the Sun to govern the supergranular properties \citep{meunier07,delmoro04}; {note, this~could prove difficult to extrapolate to other stars, where we have very little knowledge of their supergranulation properties}. Once the model star was populated with the granules and/or the supergranules, it was evolved forward in time for 12.5 years (to cover a full magnetic activity cycle) with a cadence of 30 s. For a  30~$\times$~30 Mm$^2$ region at disc center, the granulation-induced RVs were on the order of $\pm$$\sim$20~m~s$^{-1}$ (in line with the original HD simulation; note, the authors used 10 realizations of a 69 day period for this analysis, rather than the full 12.5 years simulated, but this should not impact the results since it is still much greater than the granulation timescale); this reduced by an order of magnitude when the full disc integrations were considered, with an rms over the 12.5~years of 0.8~m~s$^{-1}$ ---an order of magnitude larger than the signal from an Earth-analog.

Using these model stars, Ref.~\cite{meunier15} explored the impact of various observational strategies to maximize binning out the {solar} granulation and/or supergranulation RV variability. They varied the total observational time in a given night from $\sim$5 min up to just over 8 h, and tested the impact of spreading this time equally over a given 10 h night in intervals of 1--5 (e.g., an observing time of 2 h with 4 intervals would mean 30 min exposure times separated by 2 h each in a given night); in this way they examined the impact of numerous various exposure times. For each combination of observing time and number of observations per a night, the authors also tested the impact of binning together 1--10 consecutive nights.

\begin{wrapfigure}[25]{r}{0.5\textwidth}
\centering
\vspace{-10pt}
\includegraphics[scale=0.9]{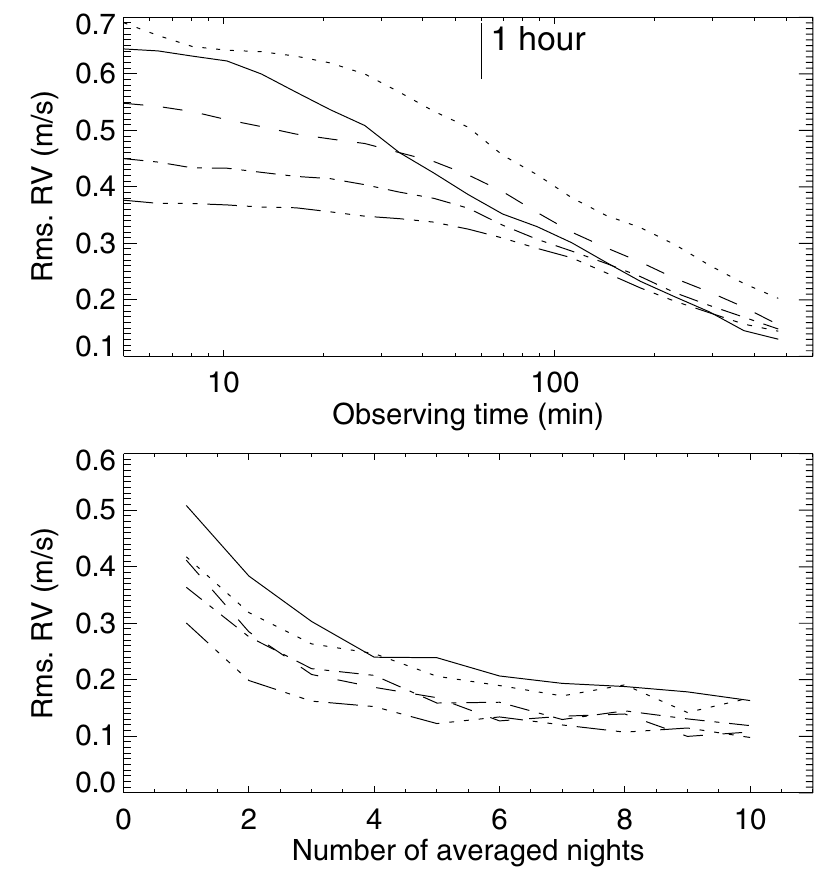}
\vspace{-5pt}
\caption{Subset of Figure~13 from \cite{meunier15} showing the residual rms after smoothing granulation noise over various amounts of observing time in a given night, with 1--5 different measurements per night (top; 1: smooth, 2: dotted, 3: dashed, \textls[-5]{4: dot-dashed, 5: dot-dot-dot-dashed). Also shown is the impact of binning all observations in 1--10 consecutive nights, if the total observing time in a night were 26 min.}}
\label{fig:meu_gran}
\end{wrapfigure}

\begin{figure}[t!]
\centering
\includegraphics[scale=1]{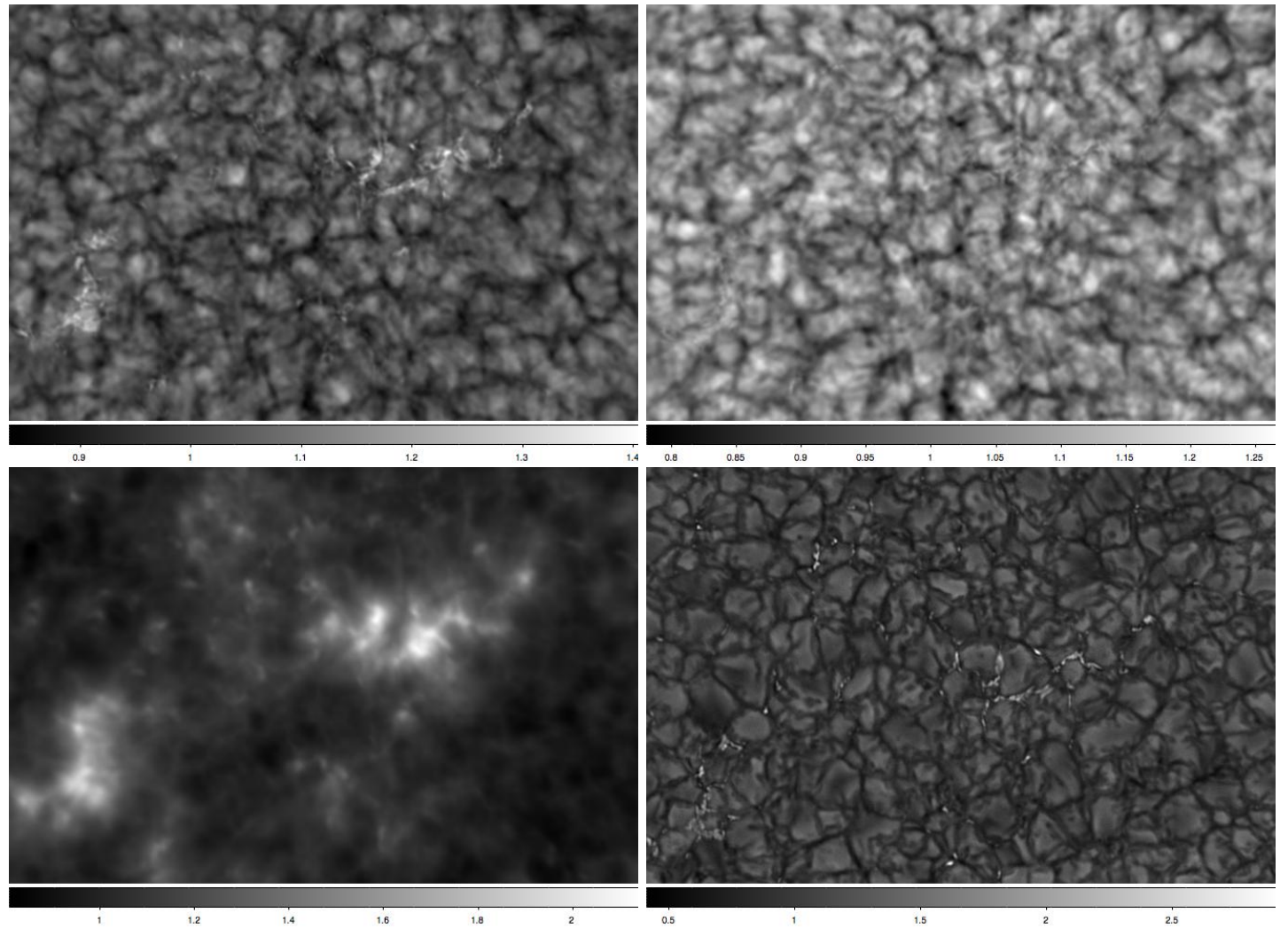}
\caption{Time-averaged solar observations in the G-band (top left), blue continuum (4170~$\angstrom$; top right), and Ca II K (bottom left), with a physical size of $\sim$61''~$\times$~57'' ($\sim$44~$\times$~42 Mm$^2$). A single G-band exposure is also shown (bottom right) for comparison, the physical size of the image is slighter larger at $\sim$67''~$\times$~65'' ($\sim$49~$\times$~47~Mm$^2$). Each time-average image was averaged over a $\sim$1 h time frame and normalized to the time-dependent mean (intensity scale displayed at the bottom of each image). Observations were obtained from the ROSA instrument on the Dunn Solar Telescope in May 2009~{\citep{jess10}}. The G-band and continuum observations have a cadence of 30.3 frames per second (0.033 s) and a~cadence of 0.528 s after speckle-reconstructed (i.e., diffraction limited) images (of 15 ms exposures) were created; the time-average images were then created by averaging all 6079 individual reconstructed images together (over a total timespan of 53 min). The Ca II K observations have a cadence of 3.8 frames per second and a cadence of 4.224 s after speckle-reconstruction (of 200 ms exposures); the time-average image was created by averaging all 761 individual reconstructed images together (over the same 53 min observing time). Although the granular lifetime is 5 - 10 min, prominent underlying structure is clearly still evident, in both the G-band and continuum images, after averaging over a 1 hr period. It is also interesting to note that the Ca II K image also still displays cellular structure from the presence of granulation, and that the regions associated with the magnetic bright points (visible in the G-band) display the most evident structure (though this may just be a byproduct of the fact that magnetic structures are bright in the Ca II K, and therefore easier to discern).}
\label{fig:jess_bin}
\end{figure}

For granulation alone, as seen in Figure~\ref{fig:meu_gran}, Ref.~\cite{meunier15} found that even after smoothing the data for 1~h the RV rms was still $\sim$0.4~m~s$^{-1}$---contradicting the claim from \cite{dumusque11a} that a 30 min duration is sufficient to bin out this noise source, and highlighting the importance of a modelling technique that does not rely on the previous generation of instrumental precision (and that can concretely discern between different activity sources). This agrees with solar observations that show the granulation structure is still clearly visible even after binning together 1 h of observations, as seen in Figure~\ref{fig:jess_bin}. The physical driver for this is likely because the granules tend to appear and disappear in the same locations. Although the granular velocities in \cite{meunier15} were derived from purely a hydrodynamical simulation, the underlying laws describing the birth, death, and evolution of the granules were derived from empirical solar observations that would naturally include some level of magnetic field; hence~the behavior of the magnetic flux could still be governing some of the granule behaviors in the end-state model observations.  Hence, the granulation results in a pink-noise-like signature that is not easily averaged out. In agreement with this, Ref.~\cite{meunier15} argue that it would require more than an~entire night of observational time to bin the granulation noise to the sub 10~cm~s$^{-1}$ level (also~shown in Figure~\ref{fig:meu_gran}). In the case of an observing time of 26 min spread over 5 separate measurements in a given night, these authors argue it would require 6--10 consecutive nights of binning to reach the 10~cm~s$^{-1}$ level for granulation alone. Moreover, when injecting a 1 Earth-mass planet into 12 years of data, only~planets with orbital periods shorter than 50 days could be found above the FAP, if sampling once every 8 days when the star is visible. Nonetheless, there were times when the long-period planet signals (e.g.,~300~and 480~d) were distinctly visible, albeit below the FAP; as such, Ref.~\cite{meunier15} argued the LPA {(local~power analysis)} method was better suited to determine detection limits (since this technique compares the strength of the periodogram peak to its surrounding peaks and not to all the peaks in the periodogram \citep{meunier12}), and claim that it should allow long-period, Earth-mass planets to be detectable if the data is sampled anywhere from every 1--20 days over a $\sim$12 year period.

On the other hand, Ref.~\cite{meunier15} found supergranulation on its own contributed to an overall rms of $\sim$0.3--1~m~s$^{-1}$.~Moreover, they found that smoothing the data over a given night, regardless of exposure time or number of observations, did not alter the rms beyond a few cm~s$^{-1}$; several days would need to be binned together to significantly beat down this stellar noise. Similar to the small-scale granulation, long-period ($>$12~d) Earth-mass planets could not be detected above the FAP; this time with even less distinction in the periodogram and therefore they further conclude that, even with the LPA technique, long-period, Earth-mass planets may be undetectable.  

When considering the combined signal of both granulation and supergranulation, Ref.~\cite{meunier15} argue that the best strategy is to sample each night up to 4 times, but that even after binning together 10~consecutive days of data that the rms could still be as high as 0.5~m~s$^{-1}$ (depending on the strength of the supergranulation). They claim the detection limits for Earth-mass planets, regardless of the sampling, were still limited to short orbital periods up to $\sim$40 days. Hence, even an excellent data sampling and over a decade of observational time may still preclude the detection of Earth-mass planets in the habitable zone of a Sun-like star, even for the most magnetically quiet stars. {On top of this, an intensive monitoring of this caliber, combined with the need for bright targets and a desire for magnetically quiet stars, means there will be strict constraints on the potential observable targets.}

Perhaps bridging the gap between empirically motivated and physically motivated strategies, Ref.~\cite{meunier17} argue that it may be possible to isolate stellar lines of varying depths to mitigate convection effects in the observed RVs. The guiding principle in this approach is that stellar line depth is correlated with the net convective blueshift \citep{AllendePrieto98, reiners16}, with shallower lines having greater blueshift. Ref.~\cite{meunier17} argue therein lies a linear relationship between the RVs observed with two different sets of absorption lines, with varying depths, and this relationship can be used to ascertain the contribution of the inhibition of convection from magnetic patches over an activity cycle. Moreover, they argue that this can be used to correct the RVs for the net convective blueshift variation and that doing so also corrects for some of the underlying shorter timescale convection/granulation noise. With such, they claim long-period, sub-Earth-mass planets may be detectable with future instrumentation. However, this approach hinges on analyzing the behavior of convection in a moderate to active star over the course of its magnetic activity cycle, wherein the inhibition of convection in plage/facular regions changes significantly; it is not designed to work on very quiet stars or short-term observations. There is also the added complexity that both the net convective blueshift and the overall granulation behavior are linked to the line depth. Deeper lines are primarily formed higher in the photosphere, so they experience convection differently than shallower lines due to a combination of different contrasts and velocity fields, with the deepest lines formed above the convection \cite{dravins81}. Along with this, a line profile corresponding to a granule will be deeper than the same profile formed within an intergranular lane; thus, the total line depth is dependent on the apparent granulation pattern and therefore linked to the granulation-induced RV shifts. To a lesser extent, the technique from Ref.~\cite{meunier17}  also relies on the assumption that the contribution from the photometric effect (i.e., dark spots or bright faculae/plage) remains constant among all stellar lines, and that the `quiet' photosphere does not change over a magnetic cycle---both of which may not necessarily be valid at the cm~s$^{-1}$ level.

\subsection{Physically Motivated Strategies}
The previous section shows how incredibly difficult it may be to average out the combined combination of stellar oscillations and granulation to a level sufficient for Earth-analog detection. If even a decade of observations may not be sufficient to bin out these noise sources, then it is clear we need to focus our approach. As such, an exploration of the underlying physics of each noise source may be key to discerning a strategy with potential to reach the 10~cm~s$^{-1}$ level necessary for an~Earth-twin scenario. Consequently, this section is subdivided into oscillation and convection noise reduction techniques. 

\subsubsection{Oscillations}
In a recent work by \cite{chaplin18}, they isolate the impact of stellar surface oscillations and explore how fine-tuning the exposure time to the host star parameters may help significantly reduce this signal in high precision RVs. The motivation for treating the oscillation impact separate to the convection lies in their inherent frequency structure. For example, most of the power for granulation phenomena reside in a much lower frequency regime and therefore a simple low-pass filter (e.g., from a finite exposure duration) will not be sufficient to fully remove this noise source. This is one of the underlying physical drivers for why \cite{dumusque11a,meunier15} found increasing the exposure time had little impact on reducing the RV variability in their simulations (which included granulation). Moreover, this is fundamentally different to stellar oscillations, which give rise to modes in a relatively narrow passband at a slightly higher frequency space---making them amenable to low-pass filtering, even for a variety of spectral types and evolutionary states. 

To simulate their model observations, Ref.~\cite{chaplin18} constructed p-mode oscillation spectra to match the Sun and other solar-type stars. In particular, they specifically calculated radial and non-radial mode-amplitudes and relative visibilities as would be expected from typical exoplanet-hunting spectrographs (this is in marked contrast to the solar, Sun-as-a-star, instruments BISON and GOLF that observe in narrow passbands; \cite{christensen-dalsgaard89,kjeldsen08, basu17}). Stars show many detectable overtones of solar-like oscillations, with the most prominent modes centered on $\nu_{\rm{max}}$.

\begin{wrapfigure}[17]{l}{0.56\textwidth}
\vspace{-10pt}%
\centering
\includegraphics[scale=1.12]{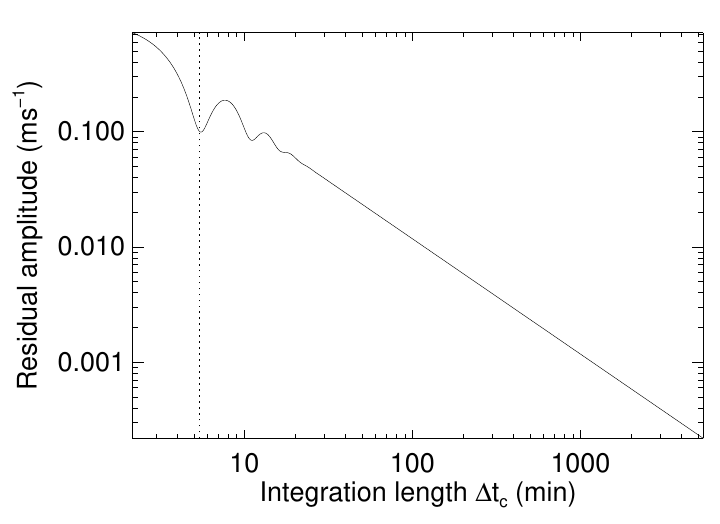}
\vspace{-13pt}
\caption{A subset of Figure~3 from \cite{chaplin18}, showing the residual RV amplitude of solar oscillations as a~function of integration (i.e., exposure) time. A vertical dashed line shows the exposure length equal to 1/$\nu_{\rm{max}}$, where $\nu_{\rm{max}}$ is the peak frequency in the envelope of solar p-modes (and therefore sensitive to the specific stellar parameters).}
\vspace{15pt}
 \label{fig:chap_sun}
\end{wrapfigure}

Once each simulated oscillation spectrum is calculated, then \cite{chaplin18} multiple this in the frequency domain by the transfer function representative of a given exposure length and examine the remaining/residual amplitudes. For an exposure of a finite duration, the transfer function can be represented by the well-known boxcar filter (\citep{chaplin14}, and references therein). Figure~\ref{fig:chap_sun} shows an example of the residual amplitudes after a variety of exposure times have been applied to solar oscillations.

Two main conclusions can be drawn from the behavior observed in Figure~\ref{fig:chap_sun}. First, that despite their larger amplitude (in comparison with granulation), for a Sun-like star, the oscillations should be easily averaged out to the 10~cm~s$^{-1}$ level or better. Second, that the behavior of the residual amplitudes is tied to the stellar parameters at a level significant for the next generation of spectrographs; simply lengthening the exposure duration does not always result in a significant noise reduction, and~sometimes may even increase the noise level compared to shorter durations. From the solar example in Figure~\ref{fig:chap_sun}, we can see that an exposure of $\sim$5.4 min (equal to 1/$\nu_{\rm{max}}$) would result in oscillation noise on the 10~cm~s$^{-1}$ level, but increasing the exposure to $\sim$ 8 min actually doubles that noise level, and further increasing to $\sim$16.5 min only reduces the noise by $\sim$1~cm~s$^{-1}$.

\begin{figure}[H]
\centering
\includegraphics[scale=1.3]{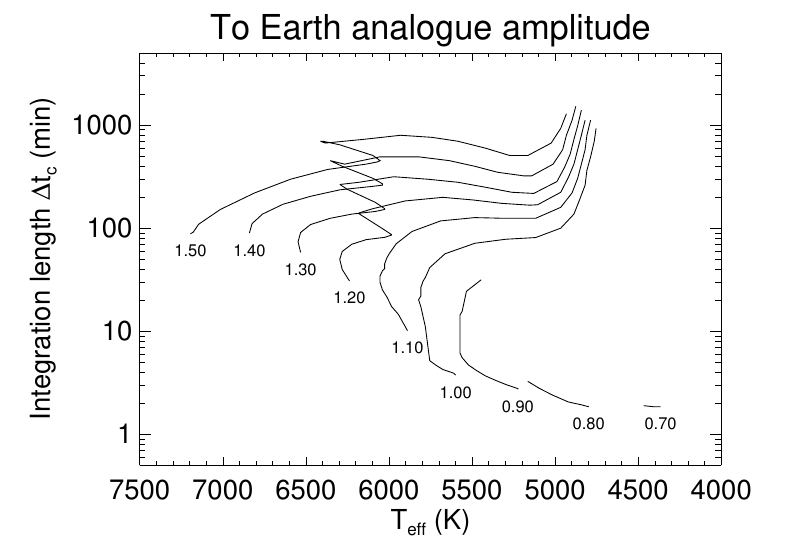}
\vspace{-8pt}
\caption{A subset of Figure~5 from \cite{chaplin18}, showing the exposure lengths required to reduce oscillation noise to a level sufficient for the detection of an Earth-mass planet in the habitable zone of stars along various evolutionary tracks, from 0.7--1.5 M$_{\odot}$.} 
 \label{fig:chap_hr}
\end{figure}

As stated above, since the stellar oscillations are tied to the host star parameters, so too are the exposure lengths necessary to bin them down to levels suitable for hunting low-mass, long-period planets. Figure~\ref{fig:chap_hr} shows the exposure times that \cite{chaplin18} argue are required to reduce the oscillation-induced noise to a level sufficient for the detection/confirmation of an Earth-mass planet in the habitable zone of stars with varying effective temperatures, surface gravities, and luminosities. These results can be replicated and/or fine-tuned to particular stars using the publicly available python code \textit{OscFilter}, which can be downloaded from: \url{https://github.com/grd349/ChaplinFilter}. Stars with lower mass and cooler effective temperatures, such as K dwarfs, may only require a few minutes to sufficiently reduce their oscillation noise; while, hotter and/or more evolved stars may need integration times much greater than 100 min to reach detection levels for Earth-mass planets in the habitable zone. {For~evolved stars that require long integration times, it may be more efficient to finely tune the observing strategy such that multiple observations in a given night are optimally spaced to beat down the oscillation noise \citep{medina18}}. As the next generation of spectrographs continue to come online, it is clear that exposure times should be tailored to the host star parameters to achieve optimal RV precisions without wasting precious telescope time. Moreover, if treated properly, stellar oscillations should not be the limiting factor in the future search for Earth-twins.

\subsubsection{Magnetoconvection}
In contrast to \cite{chaplin18}, the series by \cite{cegla13, cegla15, cegla18a, cegla18b} focuses on the impact of magnetoconvection alone; their~hypothesis is that the convection-induced line profile asymmetries responsible for this noise source can also be used to diagnose and disentangle it from planetary signals.~State-of-the-art three-dimensional magnetohydrodynamic (MHD) simulations form the backbone of the model Sun-as-a-star observations in \cite{cegla18b}; similar to \cite{meunier15}, the simulation box size is too small to permit supergranulation, as such the authors focus on the small-scale granulation. These 3D MHD simulations are coupled with 1D radiative transport to synthesis an absorption line profile for each granulation snapshot. To overcome the computational demands required to create numerous observations, tiled with realistic, independent granulation patterns, the simulation output was parameterized; this was done both at disc center \cite{cegla13} and across the stellar disc \cite{cegla18a}.

The underlying 3D MHD simulation was produced using the MURaM code \citep{vogler05}; it spans $\sim$100~min of physical time, with a cadence close to 30 s. The physical box size was 12~$\times$~12~Mm$^2$ (see~Figure~\ref{fig:2_80_vel}), with a depth of 2 Mm. The diagnostics from the MHD simulation were fed into the radiative transport code NICOLE \citep{NICOLE1, NICOLE2} to synthesize line profiles of the Fe~I~6302~$\angstrom$ line.~The~magnetoconvection in the 3D simulation naturally excites oscillations that these authors remove when parameterizing the signal from the granulation. The average magnetic field strength is 200~G, chosen to be sufficient to properly characterize the magnetic components of granulation, without strongly altering the convection characteristics (e.g., as seen in starspots). 

Each pixel in a given snapshot has an individual absorption line profile, with a variety of photospheric plasma parameters.~Since granules make up most of the surface area, and~are naturally both bright and non-magnetic, these authors use cuts in magnetic field and continuum intensity to separate the different physical components of granulation. This creates four categories: granules (non-magnetic and bright), (dark) non-magnetic and magnetic intergranular lanes, and~magnetic bright points (MBPs; bright because the intense magnetic field has evacuated the flux tubes and allows the observer to see deeper into the hotter, brighter photosphere). Since the granulation makes the stellar surface corrugated (as discussed in Section~\ref{sec:conv_osc}, and shown in Figures~\ref{fig:2_80_vel} and \ref{fig:jess_bin}), this~parameterization is performed at multiple center-to-limb angles \citep{cegla18a}. All profiles of a given category are binned together to create four time-average components for each limb angle; note, it is creating these time-averages that kills the oscillation signature. Then the probability distributions of the component filling factors can be used in conjunction with the four time-average component profiles to generate new line profiles, with the same fundamental convection characteristics as the computationally intensive 3D radiative MHD simulation \cite{cegla15, cegla18b}. 

Using this parameterization, Ref.~\cite{cegla18b} can create a stellar grid with each tile having an independent realization of granulation and integrate over the disc to mimic stellar observations. As such, Ref.~\cite{cegla18b} created 1000 Sun-as-a-star model observations, and searched for correlations between the line profile shape and the net RV. Each model observation was assumed to have been separated by at least one granulation turnover, such that each instance is independent. Each observation also represented an instantaneous moment in time; this ignores any averaging over a finite exposure length, but since granulation patterns are known to be clearly discernible over long exposures (e.g.,~see~Figure~\ref{fig:jess_bin}) this is unlikely to severely impact the results. They found many diagnostics derived from the line profile shape correlated strongly with the convective-induced RV shifts. It is important to note that the total granulation-induced RV rms for these model stars is only $\sim$10~cm~s$^{-1}$, which is {likely 3--4~times lower than} the expected variations from solar observations {\cite{elsworth94,palle99}}. This lower variability could originate from several sources: the average magnetic field strength is slightly higher than the quiet Sun, each tile is independent, each observation is instantaneous and independent, the time-series from the MHD could be under-sampled etc. For these reasons, Ref.~\cite{cegla18b} quote the fractional reduction in the RV rms as this should give an idea of how much telescope time could be saved by not needing to average as heavily and should scale with the true RV rms if the fundamental physics is correct. For~example, a 50\% reduction RV rms could mean four times less telescope time would be required to reach the same precision level; hence, even moderate noise reductions can have a significant impact.

\begin{figure}[t]
\centering
\includegraphics[trim= 2.3cm 1.1cm 0cm .1cm, clip, scale=.41]{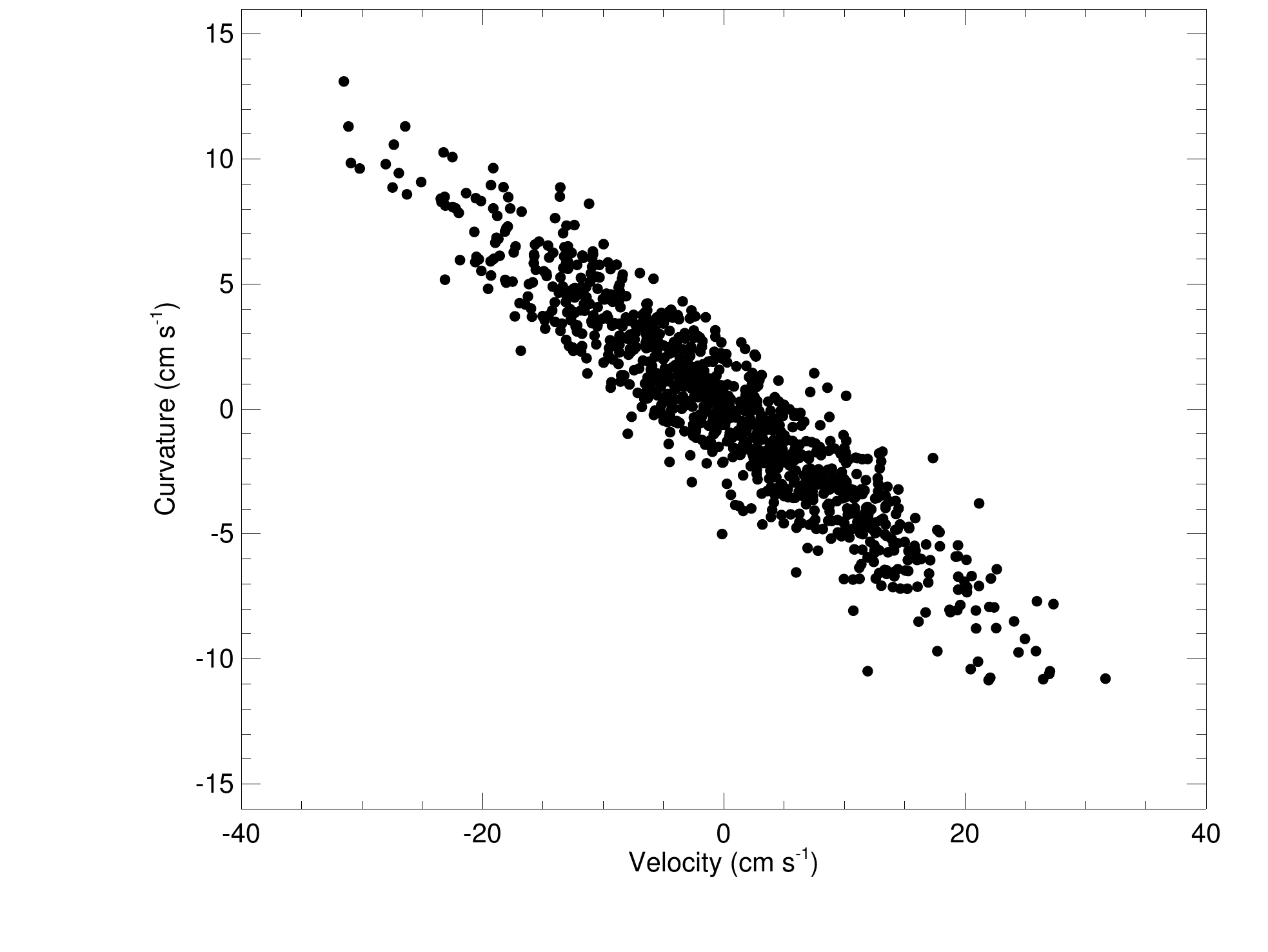}
\includegraphics[trim= 1.4cm 1.1cm .7cm .1cm, clip, scale=.41]{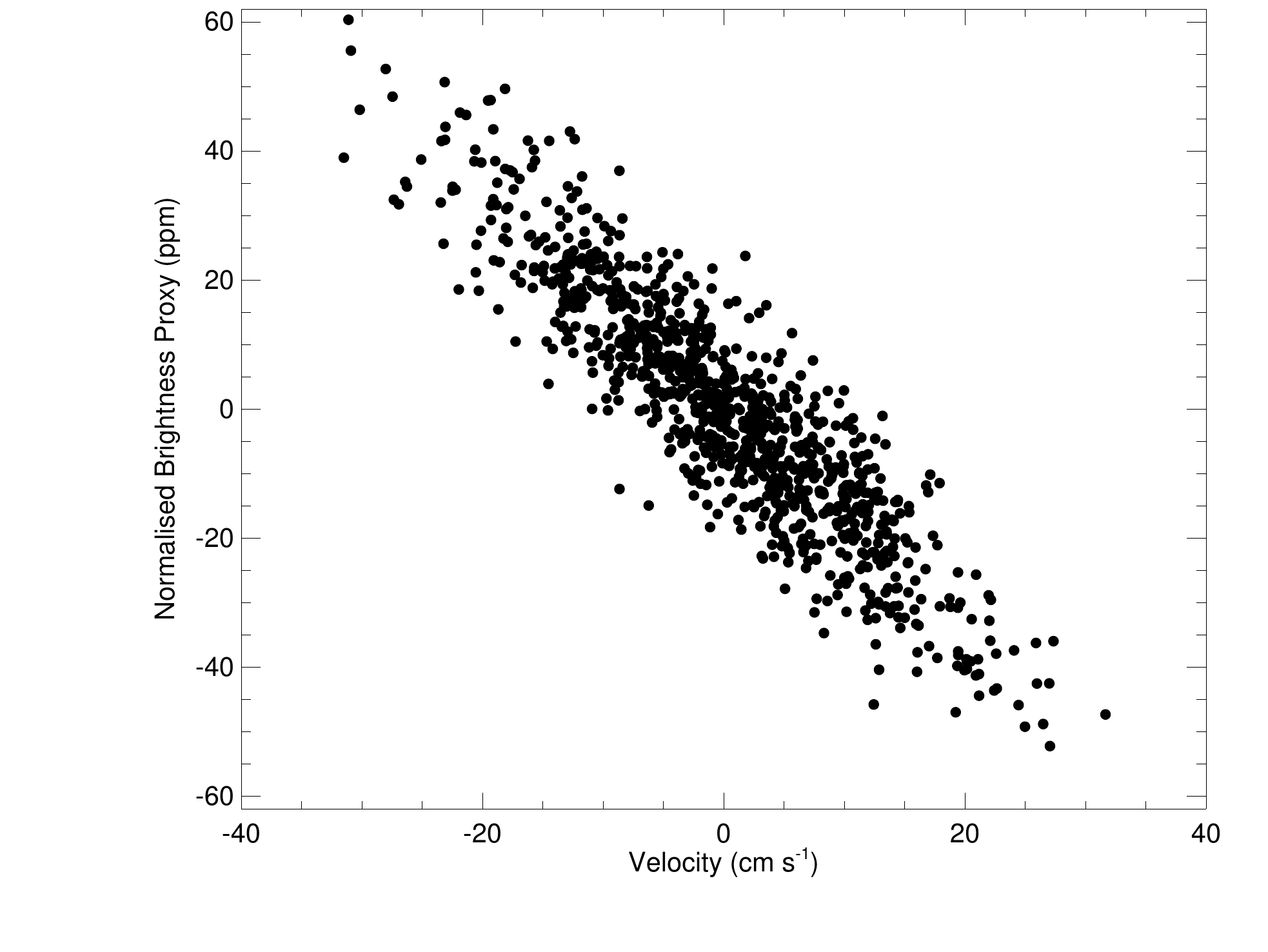}
\caption{The (mean-subtracted) bisector curvature (left) and normalized brightness proxy (right) as a function of granulation-induced RV for the Fe~I 6302$\angstrom$  line from the model star simulations of \cite{cegla18b}; the brightness was approximated by integrating the area under the line profile. A strong linear correlation is clearly discernible but would require the next generation of spectrographs and space-based photometric missions to be empirically observed.}
\label{fig:ceg_corr}
\end{figure}

In particular, Ref.~\cite{cegla18b} found measurements of the bisector shape \citep{queloz01,povich01,dall06} varied linearly with the induced RV, see Figure~\ref{fig:ceg_corr}, and that removing this correlation could reduce the RV rms by $\sim$50\% or more; note, if dividing the bisector into two or three segments (as done in the bisector span and bisector curvature) these ranges had to be fine-tuned to the particular `C'-shape of the observed bisector for this stellar line. Diagnostics that used information from the entire line profile had some of the strongest correlations (enabling noise reduction of $\sim$55--60\%) and were the most robust against the impact of instrumental resolution (a decrease resolution smooths out the asymmetries in the observed line profile); these included the equivalent width and $V_{asy}$, which compares the profile gradient of the red wing to the blue wing \citep{figueira15, lanza18}. In addition, they integrated the area under the line profile and used this as a proxy for photometric brightness; in contrast to \cite{meunier15}, they found a strong correlation, with the largest blueshifts occurring when the model star was brightest (also shown in Figure~\ref{fig:ceg_corr}). On~one hand, such a correlation is expected as a larger granule filling factor should lead to a brighter star, with~a~larger net blueshift. On the other hand, Ref.~\cite{meunier15} argue that such a correlation is broken by the stochastic nature of the granular evolution and the noisy relationship between observed granule size, velocity, and photometric properties. However, Ref.~\cite{cegla18b} argue that the empirical relationships used to create the model stars in \cite{meunier15} could potentially appear noisier than reality due to instrumental errors that are avoided when using a pure MHD-based background; that said, further analysis is required to determine the true nature of this relationship. If confirmed, then space-based simultaneous photometry could provide a very promising avenue to disentangle granulation-induced RV variability. Of course, it is important to keep in mind that \cite{cegla18b} only examined the effects on one stellar line profile, for one spectral type, assumed a constant average magnetic field for each tile on their model star, and ignored other effects such as supergranulation and oscillations. Regardless, this shows a clear indication of the potential power of the information contained in the spectral lines, and how we can use this to diagnose and disentangle stellar variability in an efficient manner. 

\section{Towards the Future}
\label{sec:future}
It is clear that magnetically active stars produce astrophysical noise much greater than the 0.5~m~s$^{-1}$ precision offered by the current world-leading spectrographs, and therefore that this noise must be removed for these instruments to reach their true potential. However, even for magnetically active stars, an understanding of magnetoconvection is key to disentangling active region signatures from planet-induced shifts, since it is the inhibition of convection that drives these noise sources. Moreover, even the most magnetically quiet Sun-like stars will still have a convective envelop and therefore exhibit RV shifts on the 0.1--1~m~s$^{-1}$ from the granulation and p-mode oscillations. As a result, the next generation of spectrographs coming online now (e.g., ESPRESSO), promising precisions of 10~cm~s$^{-1}$, will demand an understanding of these low-amplitude noise sources to deliver to their full capability. This is particularly critical for the search for life in the universe, as the only place we know for sure that can harbor life {is the Earth, which induces a signal with a mere} $\sim$9~cm~s$^{-1}$ amplitude. 

Previous studies have shown us that even the most optimal observing strategies may not be able to find such Earth-analogs in a typical survey duration of $\sim$4 years \cite{dumusque11a}. In fact, it could take more than a decade of observations to reach such detection thresholds with current strategies~\cite{meunier15}, and even then, such confirmations may be debatable if the planet signal falls below the FAP. However, a detailed understanding of the underlying stellar surface phenomena may allow us to tease out such minute planetary signals from among the larger stellar signals. In particular, fine-tuning the exposure lengths to the stellar parameters (i.e., surface gravity, luminosity, and effective temperature) may allow us to remove the oscillation-induced noise to levels sufficient for Earth-mass planet detections in the habitable zones of Sun-like stars \cite{chaplin18}.~Moreover, even though it may not be possible to efficiently bin out the granulation noise, Ref.~\cite{cegla18b} have shown that it may be possible to disentangle it by analyzing its imprint on the stellar absorption line shapes; even if such techniques cannot remove all of the granulation noise, they may help significantly reduce the overall observation time necessary to confirm Earth-like planets---making such confirmations more observationally feasible.~Nonetheless, it is important to note that while the results of \cite{cegla18b} promise great potential for granulation noise mitigation, they need to be expanded to a variety of stellar lines, magnetic field strengths, and spectral types etc. Additionally, if the convection cannot be completely disentangled in the RVs, it may be important for detection techniques to try to differentiate this colored-noise from planetary signals---e.g., see \cite{sulis17,sulis17b} for how we may be able to use HD/MHD simulations to train and standardize periodograms to counteract the convection noise and appropriately attribute false alarm and planet detection probabilities amid this noise.

Naturally, such theoretically motivated noise reduction approaches need to be validated empirically with both solar and next generation stellar spectrographs. Consequently, it is a combination of empirical and physically motivated strategies that is likely the key to overcoming the barriers of astrophysical noise. This last aspect may be particularly pertinent at the 10~cm~s$^{-1}$ regime, which will open the door to many other low-amplitude noise sources (e.g., meridional flows \cite{beckers07,makarov10} or variable gravitational redshift \citep{cegla12}), and may bring with it a host of unforeseen stellar noise sources. Thus, it is clear from the current literature that an understanding of the stellar hosts is crucial to the future confirmation and characterization of long-period (habitable{-zone}), Earth-mass planets.

\vspace{6pt}

\funding{The author acknowledges financial support from the National Centre for Competence in Research (NCCR) PlanetS, supported by the Swiss National Science Foundation (SNSF).}

\acknowledgments{The author would like to thank {the three anonymous referees and} A. Su\'arez Mascare{\~n}o for careful readings and constructive comments on previous versions of this manuscript, as well as X. Dumusque, N.~Meunier, W. J. Chaplin and S. Sulis for their very helpful feedback and clarifications. Additional thanks to \mbox{D. F. Gray}, J. L{\"o}hner-B{\"o}ttcher, K. Strassmeier, X. Dumusque, N. Meunier. D. Jess,  and W. J. Chaplin for the permission to reproduce their figures/images in this review. This work has made use of NASA’s Astrophysics Data System Bibliographic Services.}

\conflictsofinterest{The author declares no conflict of interest.} 


\begin{thebibliography}{999}
\bibitem[{Pepe} \em{et~al.}(2014){Pepe}, {Molaro}, {Cristiani}, {Rebolo},
  {Santos}, {Dekker}, {M{\'e}gevand}, {Zerbi}, {Cabral}, {Di Marcantonio},
  {Abreu}, {Affolter}, {Aliverti}, {Allende Prieto}, {Amate}, {Avila},
  {Baldini}, {Bristow}, {Broeg}, {Cirami}, {Coelho}, {Conconi}, {Coretti},
  {Cupani}, {D'Odorico}, {De Caprio}, {Delabre}, {Dorn}, {Figueira}, {Fragoso},
  {Galeotta}, {Genolet}, {Gomes}, {Gonz{\'a}lez Hern{\'a}ndez}, {Hughes},
  {Iwert}, {Kerber}, {Landoni}, {Lizon}, {Lovis}, {Maire}, {Mannetta},
  {Martins}, {Monteiro}, {Oliveira}, {Poretti}, {Rasilla}, {Riva}, {Santana
  Tschudi}, {Santos}, {Sosnowska}, {Sousa}, {Span{\'o}}, {Tenegi}, {Toso},
  {Vanzella}, {Viel}, and {Zapatero Osorio}]{pepe14}
{Pepe}, F.; {Molaro}, P.; {Cristiani}, S.; {Rebolo}, R.; {Santos}, N.C.;
  {Dekker}, H.; {M{\'e}gevand}, D.; {Zerbi}, F.M.; {Cabral}, A.; {Di
  Marcantonio}, P.; et al.
\newblock {ESPRESSO: The next European exoplanet hunter}.
\newblock {\em Astron. Nachr} {\bf 2014}, {\em 335},~8,
\newblock
  doi:12.1002/asna.201312004.

\bibitem[{Gonz{\'a}lez Hern{\'a}ndez} \em{et~al.}(2017){Gonz{\'a}lez
  Hern{\'a}ndez}, {Pepe}, {Molaro}, and {Santos}]{gonzalez17}
{Gonz{\'a}lez Hern{\'a}ndez}, J.I.; {Pepe}, F.; {Molaro}, P.; {Santos}, N.
  {ESPRESSO on VLT: An Instrument for Exoplanet Research}.
\newblock In {\em Handbook of Exoplanets};  Springer: Cham, Switzerland, {\bf 2017};
\newblock
  doi:10.1007/978-3-319-30648-3-157-1.


\bibitem[{Jurgenson}, C. and {Fischer}, D. and {McCracken}, T. and {Sawyer}, D. and 
	{Szymkowiak}, A. and {Davis}, A. and {Muller}, G. and {Santoro}, F.
	]{jurgenson16}
	{Jurgenson}, C.; {Fischer}, D.; {McCracken}, T.; {Sawyer}, D.; 
	{Szymkowiak}, A.; {Davis}, A.; {Muller}, G.; {Santoro}, F.
\newblock  {\em EXPRES: a next generation RV spectrograph in the search for earth-like worlds}
\newblock  In {\em Ground-based and Airborne Instrumentation for Astronomy VI}; International Society for Optics and Photonics: San Diego, CA, USA, {\bf 2016}
\newblock
doi:10.1117/12.2233002

\bibitem[{Fischer}, D. and {Jurgenson}, C. and {McCracken}, T. and {Sawyer}, D. and 
	{Blackman}, R. and {Szymkowiak}, A.~E.]{fischer17}
	{Fischer}, D.; {Jurgenson}, C.; {McCracken}, T.; {Sawyer}, D.; 
	{Blackman}, R.; {Szymkowiak}, A.~E
\newblock {EXPRES: the EXtreme PREcision Spectrograph at the Discovery Channel Telescope}.
\newblock {\em American Astronomical Society Meeting Abstracts \#229} {\bf 2017},
  http://adsabs.harvard.edu/abs/2017AAS...22912604F


\bibitem[{Figueira}(2018)]{figueira18}
{Figueira}, P.
\newblock {Deriving High-Precision Radial Velocities}.
\newblock In {\em Asteroseismology and Exoplanets: Listening to the Stars and
  Searching for New Worlds}; Springer International Publishing AG: Basel, Switzerland, Volume {49}, p. 181, {\bf 2018}
\newblock
  doi:10.1007/978-3-319-59315-9\_10.

\bibitem[{Saar} and {Donahue}(1997)]{saar97}
{Saar}, S.H.; {Donahue}, R.A.
\newblock {Activity-related Radial Velocity Variation in Cool Stars}. \emph{Astrophys. J.} {\bf
  1997}.
\newblock {\em 485},~319--327,
\newblock
  doi:10.1086/304392.

\bibitem[{Schrijver} and {Zwaan}(2000)]{schrijver00}
\textls[-15]{{Schrijver}, C.J.; {Zwaan}, C.
\newblock {\em {Solar and Stellar Magnetic Activity}}; Cambridge University
  Press: Cambridge, UK, 2000.}

\bibitem[{Hatzes}(2002)]{hatzes02}
{Hatzes}, A.P.
\newblock {Starspots and exoplanets}.
\newblock {\em Astron.~Nachr.} {\bf 2002}, {\em 323},~392--394,
\newblock
  doi:10.1002/1521-3994(200208)323:3/4<392::AID-ASNA392>3.0.CO;2-M.

\bibitem[{Bonfils} \em{et~al.}(2007){Bonfils}, {Mayor}, {Delfosse},
  {Forveille}, {Gillon}, {Perrier}, {Udry}, {Bouchy}, {Lovis}, {Pepe},
  {Queloz}, {Santos}, and {Bertaux}]{bonfils07}
{Bonfils}, X.; {Mayor}, M.; {Delfosse}, X.; {Forveille}, T.; {Gillon}, M.;
  {Perrier}, C.; {Udry}, S.; {Bouchy}, F.; {Lovis},~C.; {Pepe}, F.; et al.
\newblock {The HARPS search for southern extra-solar planets. X. A m sin i = 11
  M$\_{\oplus}$ planet around the nearby spotted M dwarf <ASTROBJ>GJ
  674</ASTROBJ>}.
\newblock {\em Astron. Astrophys.} {\bf 2007}, {\em 474},~293--299,
\newblock
  doi:10.1051/0004-6361:20077068.

\bibitem[{Desort} \em{et~al.}(2007){Desort}, {Lagrange}, {Galland}, {Udry}, and
  {Mayor}]{desort07}
{Desort}, M.; {Lagrange}, A.M.; {Galland}, F.; {Udry}, S.; {Mayor}, M.
\newblock {Search for exoplanets with the radial-velocity technique:
  Quantitative diagnostics of stellar activity}.
\newblock {\em Astron. Astrophys.} {\bf 2007}, {\em 473},~983--993,
\newblock
  doi:10.1051/0004-6361:20078144.

\bibitem[{Melo} \em{et~al.}(2007){Melo}, {Santos}, {Gieren}, {Pietrzynski},
  {Ruiz}, {Sousa}, {Bouchy}, {Lovis}, {Mayor}, {Pepe}, {Queloz}, {da Silva},
  and {Udry}]{melo07}
{Melo}, C.; {Santos}, N.C.; {Gieren}, W.; {Pietrzynski}, G.; {Ruiz}, M.T.;
  {Sousa}, S.G.; {Bouchy}, F.; {Lovis}, C.; {Mayor},~M.; {Pepe}, F.; et al.
\newblock {A new Neptune-mass planet orbiting HD 219828}.
\newblock {\em Astron. Astrophys.} {\bf 2007}, {\em 467},~721--727,
\newblock
  doi:10.1051/0004-6361:20066845.

\bibitem[{Queloz} \em{et~al.}(2009){Queloz}, {Bouchy}, {Moutou},
  et~al.]{queloz09}
{Queloz}, D.; {Bouchy}, F.; {Moutou}, C.; Hatzes, A.; Hébrard, G.; Alonso, R.; Auvergne, M.; Baglin, A.; Barbieri,~M.; Barge, P.; et al.
\newblock {The CoRoT-7 planetary system: Two orbiting super-Earths}.
\newblock {\em Astron. Astrophys.} {\bf 2009}, {\em 506},~303--319,
\newblock
  doi:10.1051/0004-6361/200913096.

\bibitem[{Dumusque} \em{et~al.}(2011){Dumusque}, {Santos}, {Udry}, {Lovis}, and
  {Bonfils}]{dumusque11b}
{Dumusque}, X.; {Santos}, N.C.; {Udry}, S.; {Lovis}, C.; {Bonfils}, X.
\newblock {Planetary detection limits taking into account stellar noise. II.
  Effect of stellar spot groups on radial-velocities}.
\newblock {\em Astron. Astrophys.} {\bf 2011}, {\em 527},~A82,
\newblock
  doi:10.1051/0004-6361/201015877.

\bibitem[{Aigrain} \em{et~al.}(2012){Aigrain}, {Pont}, and {Zucker}]{aigrain12}
{Aigrain}, S.; {Pont}, F.; {Zucker}, S.
\newblock {A simple method to estimate radial velocity variations due to
  stellar activity using photometry}.
\newblock {\em Mon. Not. R. Astron. Soc.} {\bf 2012}, {\em 419},~3147--3158,
\newblock
  doi:10.1111/j.1365-2966.2011.19960.x.

\bibitem[{Boisse} \em{et~al.}(2009){Boisse}, {Moutou}, {Vidal-Madjar},
  et~al.]{boisse09}
{Boisse}, I.; {Moutou}, C.; {Vidal-Madjar}, A.; Bouchy, F.; Pont, F.; Hébrard, G.; Bonfils, X.; Croll, B.; Delfosse, X.; Desort, M.; et al.
\newblock {Stellar activity of planetary host star HD 189 733}.
\newblock {\em Astron. Astrophys.} {\bf 2009}, {\em 495},~959--966,
\newblock
  doi:10.1051/0004-6361:200810648.

\bibitem[{Boisse} \em{et~al.}(2011){Boisse}, {Bouchy}, {H{\'e}brard},
  {Bonfils}, {Santos}, and {Vauclair}]{boisse11}
{Boisse}, I.; {Bouchy}, F.; {H{\'e}brard}, G.; {Bonfils}, X.; {Santos}, N.;
  {Vauclair}, S.
\newblock {Disentangling between stellar activity and planetary signals}.
\newblock {\em Astron. Astrophys.} {\bf 2011}, {\em 528},~A4,
\newblock
  doi:10.1051/0004-6361/201014354.

\bibitem[{Dumusque} \em{et~al.}(2014){Dumusque}, {Boisse}, and
  {Santos}]{dumusque14b}
{Dumusque}, X.; {Boisse}, I.; {Santos}, N.C.
\newblock {SOAP 2.0: A Tool to Estimate the Photometric and Radial Velocity
  Variations Induced by Stellar Spots and Plages}.
\newblock {\em Astrophys. J.} {\bf 2014}, {\em 796},~132,
\newblock
  doi:10.1088/0004-637X/796/2/132.

\bibitem[{Hatzes} \em{et~al.}(2010){Hatzes}, {Dvorak}, {Wuchterl}, {Guterman},
  {Hartmann}, {Fridlund}, {Gandolfi}, {Guenther}, and {P{\"a}tzold}]{hatzes10}
{Hatzes}, A.P.; {Dvorak}, R.; {Wuchterl}, G.; {Guterman}, P.; {Hartmann}, M.;
  {Fridlund}, M.; {Gandolfi}, D.; {Guenther},~E.; {P{\"a}tzold}, M.
\newblock {An investigation into the radial velocity variations of CoRoT-7}.
\newblock {\em Astron. Astrophys.} {\bf 2010}, {\em 520},~A93,
\newblock
  doi:10.1051/0004-6361/201014795.

\bibitem[{Haywood} \em{et~al.}(2014){Haywood}, {Collier Cameron}, {Queloz},
  {Barros}, {Deleuil}, {Fares}, {Gillon}, {Lanza}, {Lovis}, {Moutou}, {Pepe},
  {Pollacco}, {Santerne}, {S{\'e}gransan}, and {Unruh}]{haywood14}
{Haywood}, R.D.; {Collier Cameron}, A.; {Queloz}, D.; {Barros}, S.C.C.;
  {Deleuil}, M.; {Fares}, R.; {Gillon}, M.; {Lanza},~A.F.; {Lovis}, C.;
  {Moutou}, C.; et al.
\newblock {Planets and stellar activity: Hide and seek in the CoRoT-7 system}.
\newblock {\em Mon.~Not. R. Astron. Soc.} {\bf 2014}, {\em 443},~2517--2531,
\newblock
  doi:10.1093/mnras/stu1320.

\bibitem[{Haywood} \em{et~al.}(2016){Haywood}, {Collier Cameron}, {Unruh},
  {Lovis}, {Lanza}, {Llama}, {Deleuil}, {Fares}, {Gillon}, {Moutou}, {Pepe},
  {Pollacco}, {Queloz}, and {S{\'e}gransan}]{haywood16}
{Haywood}, R.D.; {Collier Cameron}, A.; {Unruh}, Y.C.; {Lovis}, C.; {Lanza},
  A.F.; {Llama}, J.; {Deleuil}, M.; {Fares},~R.; {Gillon}, M.; {Moutou}, C.;
  et al.
\newblock {The Sun as a planet-host star: Proxies from SDO images for HARPS
  radial-velocity variations}.
\newblock {\em Mon. Not. R. Astron. Soc.} {\bf 2016}, {\em 457},~3637--3651,
\newblock
  doi:10.1093/mnras/stw187.

\bibitem[{H{\'e}brard} \em{et~al.}(2016){H{\'e}brard}, {Donati}, {Delfosse},
  {Morin}, {Moutou}, and {Boisse}]{hebrard16}
{H{\'e}brard}, {\'E}.M.; {Donati}, J.F.; {Delfosse}, X.; {Morin}, J.; {Moutou},
  C.; {Boisse}, I.
\newblock {Modelling the RV jitter of early-M dwarfs using tomographic
  imaging}.
\newblock {\em Mon. Not. R. Astron. Soc.} {\bf 2016}, {\em 461},~1465--1497,
\newblock
  doi:10.1093/mnras/stw1346.

\bibitem[{Herrero} \em{et~al.}(2016){Herrero}, {Ribas}, {Jordi}, {Morales},
  {Perger}, and {Rosich}]{herrero16}
{Herrero}, E.; {Ribas}, I.; {Jordi}, C.; {Morales}, J.C.; {Perger}, M.;
  {Rosich}, A.
\newblock {Modelling the photosphere of active stars for planet detection and
  characterization}.
\newblock {\em Astron. Astrophys.} {\bf 2016}, {\em 586},~A131,
\newblock
  doi:10.1051/0004-6361/201425369.

\bibitem[{L{\'o}pez-Morales} \em{et~al.}(2016){L{\'o}pez-Morales}, {Haywood},
  {Coughlin}, {Zeng}, {Buchhave}, {Giles}, {Affer}, {Bonomo}, {Charbonneau},
  {Collier Cameron}, {Consentino}, {Dressing}, {Dumusque}, {Figueira},
  {Fiorenzano}, {Harutyunyan}, {Johnson}, {Latham}, {Lopez}, {Lovis},
  {Malavolta}, {Mayor}, {Micela}, {Molinari}, {Mortier}, {Motalebi},
  {Nascimbeni}, {Pepe}, {Phillips}, {Piotto}, {Pollacco}, {Queloz}, {Rice},
  {Sasselov}, {Segransan}, {Sozzetti}, {Udry}, {Vanderburg}, and
  {Watson}]{lopezmorales16}
{L{\'o}pez-Morales}, M.; {Haywood}, R.D.; {Coughlin}, J.L.; {Zeng}, L.;
  {Buchhave}, L.A.; {Giles}, H.A.C.; {Affer}, L.; {Bonomo}, A.S.;
  {Charbonneau}, D.; {Collier Cameron}, A.; et al.
\newblock {Kepler-21b: A Rocky Planet Around a V = 8.25 Magnitude Star}.
\newblock {\em Astron. J.} {\bf 2016}, {\em 152},~204,
\newblock
  doi:10.3847/0004-6256/152/6/204.

\bibitem[{Meunier} \em{et~al.}(2010{\natexlab{a}}){Meunier}, {Desort}, and
  {Lagrange}]{meunier10a}
{Meunier}, N.; {Desort}, M.; {Lagrange}, A.M.
\newblock {Using the Sun to estimate Earth-like planets detection capabilities. II. Impact of plages}.
\newblock {\em Astron. Astrophys.} {\bf 2010}, {\em 512},~A39,
\newblock
  doi:10.1051/0004-6361/200913551.

\bibitem[{Meunier} \em{et~al.}(2010{\natexlab{b}}){Meunier}, {Lagrange}, and
  {Desort}]{meunier10b}
{Meunier}, N.; {Lagrange}, A.M.; {Desort}, M.
\newblock {Reconstructing the solar integrated radial velocity using MDI/SOHO}.
\newblock {\em Astron. Astrophys.} {\bf 2010}, {\em 519},~A66,
\newblock
  doi:10.1051/0004-6361/201014199.

\bibitem[{Meunier} \em{et~al.}(2017){Meunier}, {Lagrange}, and
  {Borgniet}]{meunier17}
{Meunier}, N.; {Lagrange}, A.M.; {Borgniet}, S.
\newblock {A new method of correcting radial velocity time series for
  inhomogeneous convection}.
\newblock {\em Astron. Astrophys.} {\bf 2017}, {\em 607},~A6,
\newblock
  doi:10.1051/0004-6361/201630328.

\bibitem[{Gray}(2005)]{gray05}
\textls[-20]{{Gray}, D.F.
\newblock {\em {The Observation and Analysis of Stellar Photospheres}}; Cambridge University Press: Cambridge, UK, {\bf 2005}.}

\bibitem[{Meunier} \em{et~al.}(2017{\natexlab{a}}){Meunier}, {Lagrange},
  {Mbemba Kabuiku}, {Alex}, {Mignon}, and {Borgniet}]{meunier17a}
{Meunier}, N.; {Lagrange}, A.M.; {Mbemba Kabuiku}, L.; {Alex}, M.; {Mignon},
  L.; {Borgniet}, S.
\newblock {Variability of stellar granulation and convective blueshift with
  spectral type and magnetic activity. I. K and G main sequence stars}.
\newblock {\em Astron. Astrophys.} {\bf 2017}, {\em 597},~A52,
\newblock
  doi:10.1051/0004-6361/201629052.

\bibitem[{Meunier} \em{et~al.}(2017{\natexlab{b}}){Meunier}, {Mignon}, and
  {Lagrange}]{meunier17b}
{Meunier}, N.; {Mignon}, L.; {Lagrange}, A.M.
\newblock {Variability in stellar granulation and convective blueshift with
  spectral type and magnetic activity. II. From young to old main-sequence
  K-G-F stars}.
\newblock {\em Astron. Astrophys.} {\bf 2017}, {\em 607},~A124,
\newblock
  doi:10.1051/0004-6361/201731017.

\bibitem[{Meunier} and {Lagrange}(2013)]{meunier13}
\textls[-5]{{Meunier}, N.; {Lagrange}, A.M.
\newblock {Using the Sun to estimate Earth-like planets detection capabilities.
  IV. Correcting for the convective component}.
\newblock {\em Astron. Astrophys.} {\bf 2013}, {\em 551},~A101,
\newblock
  doi:10.1051/0004-6361/201219917.}

\bibitem[{Dravins}(2008)]{dravins08}
{Dravins}, D.
\newblock {``Ultimate'' information content in solar and stellar spectra.
  Photospheric line asymmetries and wavelength shifts}.
\newblock {\em Astron. Astrophys.} {\bf 2008}, {\em 492},~199--213,
\newblock
  doi:10.1051/0004-6361:200810481.

\bibitem[{Dravins}(1982)]{dravins82}
{Dravins}, D.
\newblock {Photospheric spectrum line asymmetries and wavelength shifts}.
\newblock {\em Ann. Rev. Astron. Astrophys} {\bf 1982}, {\em 20},~61--89,
\newblock
  doi:10.1146/annurev.aa.20.090182.000425.

\bibitem[{Balthasar}(1985)]{balthasar85}
{Balthasar}, H.
\newblock {On the contribution of horizontal granular motions to observed
  limb-effect curves}.
\newblock {\em Sol. Phys.} {\bf 1985}, {\em 99},~31--38,
\newblock
  doi:10.1007/BF00157296.

\bibitem[{Asplund} \em{et~al.}(2000){Asplund}, {Nordlund}, {Trampedach},
  {Allende Prieto}, and {Stein}]{asplund00}
{Asplund}, M.; {Nordlund}, {\AA}.; {Trampedach}, R.; {Allende Prieto}, C.;
  {Stein}, R.F.
\newblock {Line formation in solar granulation. I. Fe line shapes, shifts and
  asymmetries}.
\newblock {\em Astron. Astrophys.} {\bf 2000}, {\em 359},~729--742.

\bibitem[{Cegla} \em{et~al.}(2018){Cegla}, {Watson}, {Shelyag}, {Chaplin},
  {Davies}, {Mathioudakis}, {Palumbo}, {Saar}, and {Haywood}]{cegla18a}
{Cegla}, H.M.; {Watson}, C.A.; {Shelyag}, S.; {Chaplin}, W.J.; {Davies}, G.R.;
  {Mathioudakis}, M.; {Palumbo},~M.L.,~III; {Saar}, S.H.; {Haywood}, R.D.
\newblock {Stellar Surface Magneto-Convection as a Source of Astrophysical
  Noise II. Center-to-Limb Parameterisation of Absorption Line Profiles and
  Comparison to Observations}.
\newblock {\em Astrophys. J.} {\bf 2018}, {\em 866},~755

\bibitem[{Shporer} and {Brown}(2011)]{shporer11}
{Shporer}, A.; {Brown}, T.
\newblock {The Impact of the Convective Blueshift Effect on Spectroscopic
  Planetary Transits}.
\newblock {\em Astrophys. J.} {\bf 2011}, {\em 733},~30,
\newblock
  doi:10.1088/0004-637X/733/1/30.

\bibitem[{Cegla} \em{et~al.}(2016){Cegla}, {Oshagh}, {Watson}, {Figueira},
  {Santos}, and {Shelyag}]{cegla16a}
\textls[-15]{{Cegla}, H.M.; {Oshagh}, M.; {Watson}, C.A.; {Figueira}, P.; {Santos}, N.C.;
  {Shelyag}, S.
\newblock {Modeling the Rossiter-McLaughlin Effect: Impact of the Convective
  Center-to-limb Variations in the Stellar Photosphere}.}
\newblock {\em Astrophys. J.} {\bf 2016}, {\em 819},~67,
\newblock
  doi:10.3847/0004-637X/819/1/67.

\bibitem[{Apai} \em{et~al.}(2018){Apai}, {Rackham}, {Giampapa}, {Angerhausen},
  {Teske}, {Barstow}, {Carone}, {Cegla}, {Domagal-Goldman}, {Espinoza},
  {Giles}, {Gully-Santiago}, {Haywood}, {Hu}, {Jordan}, {Kreidberg}, {Line},
  {Llama}, {L{\'o}pez-Morales}, {Marley}, and {de Wit}]{apai18}
{Apai}, D.; {Rackham}, B.V.; {Giampapa}, M.S.; {Angerhausen}, D.; {Teske}, J.;
  {Barstow}, J.; {Carone}, L.; {Cegla},~H.; {Domagal-Goldman}, S.D.;
  {Espinoza}, N.; et al.
\newblock {Understanding Stellar Contamination in Exoplanet Transmission
  Spectra as an Essential Step in Small Planet Characterization}.
\newblock {\em arXiv} {\bf 2018}, arXiv:1803.08708.

\bibitem[{Rieutord} and {Rincon}(2010)]{rieutord10}
\textls[-5]{{Rieutord}, M.; {Rincon}, F.
\newblock {The Sun's Supergranulation}.
\newblock {\em Living Rev. Sol. Phys.} {\bf 2010}, {\em 7},~2,
\newblock
  doi:10.12942/lrsp-2010-2.}

\bibitem[{Rincon} and {Rieutord}(2018)]{rincon18}
{Rincon}, F.; {Rieutord}, M.
\newblock {The Sun's supergranulation}.
\newblock {\em Living Rev. Sol. Phys.} {\bf 2018}, {\em 15},~6,
\newblock
  doi:10.1007/s41116-018-0013-5.

\bibitem[{Roudier} \em{et~al.}(2016){Roudier}, {Malherbe}, {Rieutord}, and
  {Frank}]{roudier16}
{Roudier}, T.; {Malherbe}, J.M.; {Rieutord}, M.; {Frank}, Z.
\newblock {Relation between trees of fragmenting granules and supergranulation
  evolution}.
\newblock {\em Astron. Astrophys.} {\bf 2016}, {\em 590},~A121,
\newblock
  doi:10.1051/0004-6361/201628111.

\bibitem[{Malherbe} \em{et~al.}(){Malherbe}, {Roudier}, {Stein}, and
  {Frank}]{malherbe18}
{Malherbe}, J.M.; {Roudier}, T.; {Stein}, R.; {Frank}, Z.
\newblock {Dynamics of Trees of Fragmenting Granules in the Quiet Sun:
  Hinode/SOT Observations Compared to Numerical Simulation}.
\newblock {\em Sol. Phys.} \textbf{2018}, \emph{293}, 4.



\bibitem[{Chaplin} and {Miglio}(2013)]{chaplin13}
{Chaplin}, W.J.; {Miglio}, A.
\newblock {Asteroseismology of Solar-Type and Red-Giant Stars}.
\newblock {\em Annu. Rev. Astron. Astrophys.} {\bf 2013}, {\em 51},~353--392,
\newblock
  doi:10.1146/annurev-astro-082812-140938.

\bibitem[{L{\"o}hner-B{\"o}ttcher} \em{et~al.}(2018){L{\"o}hner-B{\"o}ttcher},
  {Schmidt}, {Stief}, {Steinmetz}, and {Holzwarth}]{lohner18}
{L{\"o}hner-B{\"o}ttcher}, J.; {Schmidt}, W.; {Stief}, F.; {Steinmetz}, T.;
  {Holzwarth}, R.
\newblock {Convective blueshifts in the solar atmosphere. I. Absolute
  measurements with LARS of the spectral lines at 6302 {\AA}}.
\newblock {\em Astron. Astrophys.} {\bf 2018}, {\em 611},~A4,
\newblock
  doi:10.1051/0004-6361/201732107.

\bibitem[{Strassmeier} \em{et~al.}(2018){Strassmeier}, {Ilyin}, and
  {Steffen}]{strassmeier18}
{Strassmeier}, K.G.; {Ilyin}, I.; {Steffen}, M.
\newblock {PEPSI deep spectra. I. The Sun-as-a-star}.
\newblock {\em Astron. Astrophys.} {\bf 2018}, {\em 612},~A44,
\newblock
  doi:10.1051/0004-6361/201731631.

\bibitem[{Stein}(2012)]{stein12}
{Stein}, R.F.
\newblock {Solar Surface Magneto-Convection}.
\newblock {\em Living Rev. Sol. Phys.} {\bf 2012}, {\em 9},~4.

\bibitem[{Nordlund} \em{et~al.}(2009){Nordlund}, {Stein}, and
  {Asplund}]{nordlund09}
{Nordlund}, {\AA}.; {Stein}, R.F.; {Asplund}, M.
\newblock {Solar Surface Convection}.
\newblock {\em Living Rev. Sol. Phys.} {\bf 2009}, {\em 6},~2.

\bibitem[{Gizon} and {Birch}(2005)]{gizon05}
{Gizon}, L.; {Birch}, A.C.
\newblock {Local Helioseismology}.
\newblock {\em Living Rev. Sol. Phys.} {\bf 2005}, {\em 2},~6,
\newblock
  doi:10.12942/lrsp-2005-6.

\bibitem[{Basu}(2016)]{basu16}
{Basu}, S.
\newblock {Global seismology of the Sun}.
\newblock {\em Living Rev. Sol. Phys.} {\bf 2016}, {\em 13},~2,
\newblock
  doi:10.1007/s41116-016-0003-4.

\bibitem[{Santos} \em{et~al.}(2000){Santos}, {Mayor}, {Naef}, {Pepe}, {Queloz},
  {Udry}, and {Blecha}]{santos00}
{Santos}, N.C.; {Mayor}, M.; {Naef}, D.; {Pepe}, F.; {Queloz}, D.; {Udry}, S.;
  {Blecha}, A.
\newblock {The CORALIE survey for Southern extra-solar planets. IV. Intrinsic
  stellar limitations to planet searches with radial-velocity techniques}.
\newblock {\em Astron. Astrophys.} {\bf 2000}, {\em 361},~265--272.

\bibitem[{Saar} \em{et~al.}(2003){Saar}, {Hatzes}, {Cochran}, and
  {Paulson}]{saar03}
{Saar}, S.H.; {Hatzes}, A.; {Cochran}, W.; {Paulson}, D.
\newblock {Stellar Intrinsic Radial Velocity Noise: Causes and Possible Cures}.
\newblock  In \emph{The Future of Cool-Star Astrophysics, Proceedings of the 12th Cambridge Workshop on
  Cool Stars, Stellar Systems, and the Sun, Boulder, CO, USA, 30 July--3 August 2001}; {Brown}, A., {Harper}, G.M.,
  {Ayres}, T.R., Eds.;  University of Colorado: Boulder, CO, USA, {\bf 2003}; Volume~12, pp. 694--698.
  
\bibitem[{Wright}(2005)]{wright05}
{Wright}, J.T.
\newblock {Radial Velocity Jitter in Stars from the California and Carnegie
  Planet Search at Keck Observatory}.
\newblock {\em Publ. Astron. Soc. Pac.} {\bf 2005}, {\em 117},~657--664,
\newblock
  doi:10.1086/430369.

\bibitem[{Cegla} \em{et~al.}(2014){Cegla}, {Stassun}, {Watson}, {Bastien}, and
  {Pepper}]{cegla14a}
{Cegla}, H.M.; {Stassun}, K.G.; {Watson}, C.A.; {Bastien}, F.A.; {Pepper}, J.
\newblock {Estimating Stellar Radial Velocity Variability from Kepler and
  GALEX: Implications for the Radial Velocity Confirmation of Exoplanets}.
\newblock {\em Astrophys. J.} {\bf 2014}, {\em 780},~104,
\newblock
  doi:10.1088/0004-637X/780/1/104.

\bibitem[{Bastien} \em{et~al.}(2014){Bastien}, {Stassun}, {Pepper}, {Wright},
  {Aigrain}, {Basri}, {Johnson}, {Howard}, and {Walkowicz}]{bastien14}
{Bastien}, F.A.; {Stassun}, K.G.; {Pepper}, J.; {Wright}, J.T.; {Aigrain}, S.;
  {Basri}, G.; {Johnson}, J.A.; {Howard},~A.W.; {Walkowicz}, L.M.
\newblock {Radial Velocity Variations of Photometrically Quiet,
  Chromospherically Inactive Kepler Stars: A Link between RV Jitter and
  Photometric Flicker}.
\newblock {\em Astron. J.} {\bf 2014}, {\em 147},~29,
\newblock
  doi:10.1088/0004-6256/147/2/29.

\bibitem[{Marchwinski} \em{et~al.}(2015){Marchwinski}, {Mahadevan},
  {Robertson}, {Ramsey}, and {Harder}]{marchwinski15}
{Marchwinski}, R.C.; {Mahadevan}, S.; {Robertson}, P.; {Ramsey}, L.; {Harder},
  J.
\newblock {Toward Understanding Stellar Radial Velocity Jitter as a Function of
  Wavelength: The Sun as a Proxy}.
\newblock {\em Astrophys. J.} {\bf 2015}, {\em 798},~63,
\newblock
  doi:10.1088/0004-637X/798/1/63.

\bibitem[{Yu} \em{et~al.}(2018){Yu}, {Huber}, {Bedding}, and {Stello}]{yu18}
{Yu}, J.; {Huber}, D.; {Bedding}, T.R.; {Stello}, D.
\newblock {Predicting radial-velocity jitter induced by stellar oscillations
  based on Kepler data}.
\newblock {\em Mon. Not. R. Astron. Soc.} {\bf 2018}, {\em 480},~L48--L53,
\newblock
  doi:10.1093/mnrasl/sly123.

\bibitem[{Dumusque} \em{et~al.}(2011){Dumusque}, {Udry}, {Lovis}, {Santos}, and
  {Monteiro}]{dumusque11a}
{Dumusque}, X.; {Udry}, S.; {Lovis}, C.; {Santos}, N.C.; {Monteiro}, M.J.P.F.G.
\newblock {Planetary detection limits taking into account stellar noise. I.
  Observational strategies to reduce stellar oscillation and granulation
  effects}.
\newblock {\em Astron.~Astrophys.} {\bf 2011}, {\em 525},~A140,
\newblock
  doi:10.1051/0004-6361/201014097.

\bibitem[{Harvey}(1984)]{harvey84}
{Harvey}, J.W.
\newblock {\em {Probing the Depths of a Star: The Study of the Solar Oscillation from the Space}}; NASA JPL: Pasadena, CA, USA, {\bf 1984}; Volume 400,  p. 327.

\bibitem[{Andersen} \em{et~al.}(1994){Andersen}, {Leifsen}, and
  {Toutain}]{andersen94}
{Andersen}, B.N.; {Leifsen}, T.E.; {Toutain}, T.
\newblock {Solar noise simulations in irradiance}.
\newblock {\em Sol. Phys.} {\bf 1994}, {\em 152},~247--252,
\newblock
  doi:10.1007/BF01473211.

\bibitem[{Palle} \em{et~al.}(1995){Palle}, {Jimenez}, {Perez Hernandez},
  {Regulo}, {Roca Cortes}, and {Sanchez}]{palle95}
{Palle}, P.L.; {Jimenez}, A.; {Perez Hernandez}, F.; {Regulo}, C.; {Roca
  Cortes}, T.; {Sanchez}, L.
\newblock {A measurement of the background solar velocity spectrum}.
\newblock {\em Astrophys. J.} {\bf 1995}, {\em 441},~952--959,
\newblock
  doi:10.1086/175414.

\bibitem[{Kjeldsen} \em{et~al.}(2005){Kjeldsen}, {Bedding}, {Butler},
  {Christensen-Dalsgaard}, {Kiss}, {McCarthy}, {Marcy}, {Tinney}, and
  {Wright}]{kjeldsen05}
{Kjeldsen}, H.; {Bedding}, T.R.; {Butler}, R.P.; {Christensen-Dalsgaard}, J.;
  {Kiss}, L.L.; {McCarthy}, C.; {Marcy}, G.W.; {Tinney}, C.G.; {Wright}, J.T.
\newblock {Solar-like Oscillations in {$\alpha$} Centauri B}.
\newblock {\em Astrophys. J.} {\bf 2005}, {\em 635},~1281--1290,
\newblock
  doi:10.1086/497530.

\bibitem[{Arentoft} \em{et~al.}(2008){Arentoft}, {Kjeldsen}, {Bedding},
  {Bazot}, {Christensen-Dalsgaard}, {Dall}, {Karoff}, {Carrier}, {Eggenberger},
  {Sosnowska}, {Wittenmyer}, {Endl}, {Metcalfe}, {Hekker}, {Reffert}, {Butler},
  {Bruntt}, {Kiss}, {O'Toole}, {Kambe}, {Ando}, {Izumiura}, {Sato}, {Hartmann},
  {Hatzes}, {Bouchy}, {Mosser}, {Appourchaux}, {Barban}, {Berthomieu},
  {Garcia}, {Michel}, {Provost}, {Turck-Chi{\`e}ze}, {Marti{\'c}}, {Lebrun},
  {Schmitt}, {Bertaux}, {Bonanno}, {Benatti}, {Claudi}, {Cosentino}, {Leccia},
  {Frandsen}, {Brogaard}, {Glowienka}, {Grundahl}, and {Stempels}]{arentoft08}
{Arentoft}, T.; {Kjeldsen}, H.; {Bedding}, T.R.; {Bazot}, M.;
  {Christensen-Dalsgaard}, J.; {Dall}, T.H.; {Karoff}, C.; {Carrier},~F.;
  {Eggenberger}, P.; {Sosnowska}, D.; et al.
\newblock {A Multisite Campaign to Measure Solar-like Oscillations in Procyon.
  I. Observations, Data Reduction, and Slow Variations}.
\newblock {\em Astrophys. J.} {\bf 2008}, {\em 687},~1180--1190,
\newblock
  doi:10.1086/592040.

\bibitem[{Kallinger} \em{et~al.}(2014){Kallinger}, {De Ridder}, {Hekker},
  {Mathur}, {Mosser}, {Gruberbauer}, {Garc{\'{\i}}a}, {Karoff}, and
  {Ballot}]{kallinger14}
{Kallinger}, T.; {De Ridder}, J.; {Hekker}, S.; {Mathur}, S.; {Mosser}, B.;
  {Gruberbauer}, M.; {Garc{\'{\i}}a}, R.A.; {Karoff},~C.; {Ballot},~J.
\newblock {The connection between stellar granulation and oscillation as seen
  by the Kepler mission}.
\newblock {\em Astron.~Astrophys.} {\bf 2014}, {\em 570},~A41,
\newblock
  doi:10.1051/0004-6361/201424313.

\bibitem[{Endl} \em{et~al.}(2001){Endl}, {K{\"u}rster}, {Els}, {Hatzes}, and
  {Cochran}]{endl01}
{Endl}, M.; {K{\"u}rster}, M.; {Els}, S.; {Hatzes}, A.P.; {Cochran}, W.D.
\newblock {The planet search program at the ESO Coud{\'e} Echelle spectrometer.
  II. The alpha Centauri system: Limits for planetary companions}.
\newblock {\em Astron. Astrophys.} {\bf 2001}, {\em 374},~675--681,
\newblock
  doi:10.1051/0004-6361:20010723.

\bibitem[{Efron} and {Tibshirani}(1998)]{efron98}
{Efron}, B.; {Tibshirani}, R.J.
\newblock {\em {An Introduction to the Bootstrap}}: Chapman \& Hall/CRC; {CRC Press: Boca Raton, FL, USA, }{\bf 1998}.

\bibitem[{Meunier} \em{et~al.}(2015){Meunier}, {Lagrange}, {Borgniet}, and
  {Rieutord}]{meunier15}
{Meunier}, N.; {Lagrange}, A.M.; {Borgniet}, S.; {Rieutord}, M.
\newblock {Using the Sun to estimate Earth-like planet detection capabilities.
  VI. Simulation of granulation and supergranulation radial velocity and
  photometric time series}.
\newblock {\em Astron. Astrophys.} {\bf 2015}, {\em 583},~A118,
\newblock
  doi:10.1051/0004-6361/201525721.

\bibitem[{Roudier} and {Muller}(1986)]{roudier86}
{Roudier}, T.; {Muller}, R.
\newblock {Structure of the solar granulation}.
\newblock {\em Sol. Phys.} {\bf 1986}, {\em 107},~11--26,
\newblock
  doi:10.1007/BF00155337.

\bibitem[{Hirzberger} \em{et~al.}(1997){Hirzberger}, {V{\'a}zquez}, {Bonet},
  {Hanslmeier}, and {Sobotka}]{hirzberger97}
{Hirzberger}, J.; {V{\'a}zquez}, M.; {Bonet}, J.A.; {Hanslmeier}, A.;
  {Sobotka}, M.
\newblock {Time Series of Solar Granulation Images. I. Differences between
  Small and Large Granules in Quiet Regions}.
\newblock {\em Astrophys. J.} {\bf 1997}, {\em 480},~406--419,
\newblock
  doi:10.1086/303951.

\bibitem[{Hirzberger} \em{et~al.}(1999){Hirzberger}, {Bonet}, {V{\'a}zquez},
  and {Hanslmeier}]{hirzberger99}
{Hirzberger}, J.; {Bonet}, J.A.; {V{\'a}zquez}, M.; {Hanslmeier}, A.
\newblock {Time Series of Solar Granulation Images. II. Evolution of Individual
  Granules}.
\newblock {\em Astrophys. J.} {\bf 1999}, {\em 515},~441--454,
\newblock
  doi:10.1086/307018.

\bibitem[{Rieutord} \em{et~al.}(2002){Rieutord}, {Ludwig}, {Roudier},
  {Nordlund}, and {Stein}]{rieutord02}
{Rieutord}, M.; {Ludwig}, H.G.; {Roudier}, T.; {Nordlund}, A.; {Stein}, R.
\newblock {A simulation of solar convection at supergranulation scale}.
\newblock {\em Nuovo Cimento C Geophys. Space Phys. C} {\bf 2002}, {\em
  25},~523.

\bibitem[{Beeck} \em{et~al.}(2013{\natexlab{a}}){Beeck}, {Cameron}, {Reiners},
  and {Sch{\"u}ssler}]{beeck13a}
{Beeck}, B.; {Cameron}, R.H.; {Reiners}, A.; {Sch{\"u}ssler}, M.
\newblock {Three-dimensional simulations of near-surface convection in
  main-sequence stars. I. Overall structure}.
\newblock {\em Astron. Astrophys.} {\bf 2013}, {\em 558},~A48,
\newblock
  doi:10.1051/0004-6361/201321343.

\bibitem[{Beeck} \em{et~al.}(2013{\natexlab{b}}){Beeck}, {Cameron}, {Reiners},
  and {Sch{\"u}ssler}]{beeck13}
{Beeck}, B.; {Cameron}, R.H.; {Reiners}, A.; {Sch{\"u}ssler}, M.
\newblock {Three-dimensional simulations of near-surface convection in
  main-sequence stars. II. Properties of granulation and spectral lines}.
\newblock {\em Astron. Astrophys.} {\bf 2013}, {\em 558},~A49,
\newblock
  doi:10.1051/0004-6361/201321345.

\bibitem[{Beeck} \em{et~al.}(2015{\natexlab{a}}){Beeck}, {Sch{\"u}ssler},
  {Cameron}, and {Reiners}]{beeck15a}
{Beeck}, B.; {Sch{\"u}ssler}, M.; {Cameron}, R.H.; {Reiners}, A.
\newblock {Three-dimensional simulations of near-surface convection in
  main-sequence stars. III. The structure of small-scale magnetic flux
  concentrations}.
\newblock {\em Astron.~Astrophys.} {\bf 2015}, {\em 581},~A42,
\newblock
  doi:10.1051/0004-6361/201525788.

\bibitem[{Beeck} \em{et~al.}(2015{\natexlab{b}}){Beeck}, {Sch{\"u}ssler},
  {Cameron}, and {Reiners}]{beeck15b}
{Beeck}, B.; {Sch{\"u}ssler}, M.; {Cameron}, R.H.; {Reiners}, A.
\newblock {Three-dimensional simulations of near-surface convection in
  main-sequence stars. IV. Effect of small-scale magnetic flux concentrations
  on centre-to-limb variation and spectral lines}.
\newblock {\em Astron. Astrophys.} {\bf 2015}, {\em 581},~A43,
\newblock
  doi:10.1051/0004-6361/201525874.

\bibitem[{Meunier} \em{et~al.}(2007){Meunier}, {Tkaczuk}, {Roudier}, and
  {Rieutord}]{meunier07}
{Meunier}, N.; {Tkaczuk}, R.; {Roudier}, T.; {Rieutord}, M.
\newblock {Velocities and divergences as a function of supergranule size}.
\newblock {\em Astron. Astrophys.} {\bf 2007}, {\em 461},~1141--1147,
\newblock
  doi:10.1051/0004-6361:20065625.

\bibitem[{Del Moro}(2004)]{delmoro04}
{Del Moro}, D.
\newblock {Solar granulation properties derived from three different time
  series}.
\newblock {\em Astron. Astrophys.} {\bf 2004}, {\em 428},~1007--1015,
\newblock
  doi:10.1051/0004-6361:20040466.

\bibitem[{Meunier} \em{et~al.}(2012){Meunier}, {Lagrange}, and {De
  Bondt}]{meunier12}
{Meunier}, N.; {Lagrange}, A.M.; {De Bondt}, K.
\newblock {Comparison of different exoplanet mass detection limit methods using
  a sample of main-sequence intermediate-type stars}.
\newblock {\em Astron. Astrophys.} {\bf 2012}, {\em 545},~A87,
\newblock
  doi:10.1051/0004-6361/201219163.

\bibitem[{Jess} \em{et~al.}(2010){Jess}, {Mathioudakis}, {Christian},
  {Crockett}, and {Keenan}]{jess10}
{Jess}, D.B.; {Mathioudakis}, M.; {Christian}, D.J.; {Crockett}, P.J.;
  {Keenan}, F.P.
\newblock {A Study of Magnetic Bright Points in the Na I D$_{1}$ Line}.
\newblock {\em Astrophys. J. Lett.} {\bf 2010}, {\em 719},~L134--L139,
\newblock
  doi:10.1088/2041-8205/719/2/L134.

\bibitem[{Allende Prieto} and {Garcia Lopez}(1998)]{AllendePrieto98}
{Allende Prieto}, C.; {Garcia Lopez}, R.J.
\newblock {Fe i line shifts in the optical spectrum of the Sun}.
\newblock {\em Astron. Astrophys. Suppl. Ser.} {\bf 1998}, {\em 129},~41--44,
\newblock
  doi:10.1051/aas:1998173.

\bibitem[{Reiners} \em{et~al.}(2016){Reiners}, {Mrotzek}, {Lemke}, {Hinrichs},
  and {Reinsch}]{reiners16}
{Reiners}, A.; {Mrotzek}, N.; {Lemke}, U.; {Hinrichs}, J.; {Reinsch}, K.
\newblock {The IAG solar flux atlas: Accurate wavelengths and absolute
  convective blueshift in standard solar spectra}.
\newblock {\em Astron. Astrophys.} {\bf 2016}, {\em 587},~A65,
\newblock
  doi:10.1051/0004-6361/201527530.

\bibitem[{Dravins} \em{et~al.}(1981){Dravins}, {Lindegren}, and
  {Nordlund}]{dravins81}
{Dravins}, D.; {Lindegren}, L.; {Nordlund}, A.
\newblock {Solar granulation---Influence of convection on spectral line
  asymmetries and wavelength shifts}.
\newblock {\em Astron. Astrophys.} {\bf 1981}, {\em 96},~345--364.

\bibitem[{Chaplin} \em{et~al.}(2018){Chaplin}, {Cegla}, {Watson}, and
  {Davies}]{chaplin18}
{Chaplin}, W.J.; {Cegla}, H.M.; {Watson}, C.A.; {Davies}, G.R.
\newblock {Filtering solar-like oscillations for exoplanet detection in radial
  velocity observations}.
\newblock {\em Astron. J.} {\bf 2019}, {arXiv:1903.00657}.

\bibitem[{Christensen-Dalsgaard}(1989)]{christensen-dalsgaard89}
{Christensen-Dalsgaard}, J.
\newblock {The effect of rotation on whole-disc Doppler observations of solar
  oscillations}.
\newblock {\em Mon. Not. R. Astron. Soc.} {\bf 1989}, {\em 239},~977--994,
\newblock
  doi:10.1093/mnras/239.3.977.

\bibitem[{Kjeldsen} \em{et~al.}(2008){Kjeldsen}, {Bedding}, {Arentoft},
  {Butler}, {Dall}, {Karoff}, {Kiss}, {Tinney}, and {Chaplin}]{kjeldsen08}
{Kjeldsen}, H.; {Bedding}, T.R.; {Arentoft}, T.; {Butler}, R.P.; {Dall}, T.H.;
  {Karoff}, C.; {Kiss}, L.L.; {Tinney}, C.G.; {Chaplin},~W.J.
\newblock {The Amplitude of Solar Oscillations Using Stellar Techniques}.
\newblock {\em Astrophys. J.} {\bf 2008}, {\em 682},~1370--1375,
\newblock
  doi:10.1086/589142.

\bibitem[{Basu} and {Chaplin}(2017)]{basu17}
{Basu}, S.; {Chaplin}, W.J.
\newblock {\em {Asteroseismic Data Analysis: Foundations and Techniques}};
  Princenton University Press: Princeton, NJ, USA, {\bf 2017}.

\bibitem[{Chaplin} \em{et~al.}(2014){Chaplin}, {Elsworth}, {Davies},
  {Campante}, {Handberg}, {Miglio}, and {Basu}]{chaplin14}
{Chaplin}, W.J.; {Elsworth}, Y.; {Davies}, G.R.; {Campante}, T.L.; {Handberg},
  R.; {Miglio}, A.; {Basu}, S.
\newblock {Super-Nyquist asteroseismology of solar-like oscillators with Kepler
  and K2---Expanding the asteroseismic cohort at the base of the red giant
  branch}.
\newblock {\em Mon. Not. R. Astron. Soc.} {\bf 2014}, {\em 445},~946--954,
\newblock
  doi:10.1093/mnras/stu1811.

\bibitem[{Medina} \em{et~al.}(2018){Medina}, {Johnson}, {Eastman}, and
  {Cargile}]{medina18}
{Medina}, A.A.; {Johnson}, J.A.; {Eastman}, J.D.; {Cargile}, P.A.
\newblock {Techniques for Finding Close-in, Low-mass Planets around Evolved
  Intermediate-mass Stars}.
\newblock {\em Astrophys. J.} {\bf 2018}, {\em 867},~32,
\newblock
  doi:10.3847/1538-4357/aadf82.

\bibitem[{Cegla} \em{et~al.}(2013){Cegla}, {Shelyag}, {Watson}, and
  {Mathioudakis}]{cegla13}
{Cegla}, H.M.; {Shelyag}, S.; {Watson}, C.A.; {Mathioudakis}, M.
\newblock {Stellar Surface Magneto-convection as a Source of Astrophysical
  Noise. I. Multi-component Parameterization of Absorption Line Profiles}.
\newblock {\em Astrophys. J.} {\bf 2013}, {\em 763},~95,
\newblock
  doi:10.1088/0004-637X/763/2/95.

\bibitem[{Cegla} \em{et~al.}(2015){Cegla}, {Watson}, {Shelyag}, and
  {Mathioudakis}]{cegla15}
{Cegla}, H.M.; {Watson}, C.A.; {Shelyag}, S.; {Mathioudakis}, M.
\newblock {Understanding Astrophysical Noise from Stellar Surface
  Magneto-Convection}.
\newblock  In Proceedings of the 18th Cambridge Workshop on Cool Stars, Stellar Systems, and the Sun, {Proceedings of Lowell Observatory, Flagstaff, AZ, USA, 8--14 June 2014}; Cambridge
  Workshop on Cool Stars, Stellar Systems, and the Sun; 
  {van Belle}, G.T., {Harris}, H.C., Eds.;  {\bf 2015}; Volume~18, pp. 567--574.
  
\bibitem[{Cegla} \em{et~al.}(2018){Cegla}, {Watson}, {Shelyag}, {Mathioudakis},
  and {Moutari}]{cegla18b}
{Cegla}, H.M.; {Watson}, C.A.; {Shelyag}, S.; {Mathioudakis}, M.; {Moutari}, S.
\newblock {Stellar Surface Magneto-Convection as a Source of Astrophysical
  Noise III. Sun-as-a-star Simulations and Optimal Noise Diagnostics}.
\newblock {\em Astrophys.~J.} {\bf 2019}, {arXiv:1903.08446}.

\bibitem[{V{\"o}gler} \em{et~al.}(2005){V{\"o}gler}, {Shelyag},
  {Sch{\"u}ssler}, {Cattaneo}, {Emonet}, and {Linde}]{vogler05}
{V{\"o}gler}, A.; {Shelyag}, S.; {Sch{\"u}ssler}, M.; {Cattaneo}, F.; {Emonet},
  T.; {Linde}, T.
\newblock {Simulations of magneto-convection in the solar photosphere.
  Equations, methods, and results of the MURaM code}.
\newblock {\em Astron. Astrophys.} {\bf 2005}, {\em 429},~335--351,
\newblock
  doi:10.1051/0004-6361:20041507.

\bibitem[{Socas-Navarro}(2015)]{NICOLE1}
{Socas-Navarro}, H.
\newblock {\emph{NICOLE: NLTE Stokes Synthesis/Inversion Code}};
\newblock Astrophysics Source Code Library: record ascl:1508.002  {\bf 2015}.

\bibitem[{Socas-Navarro} \em{et~al.}(2015){Socas-Navarro}, {de la Cruz
  Rodr{\'{\i}}guez}, {Asensio Ramos}, {Trujillo Bueno}, and {Ruiz
  Cobo}]{NICOLE2}
{Socas-Navarro}, H.; {de la Cruz Rodr{\'{\i}}guez}, J.; {Asensio Ramos}, A.;
  {Trujillo Bueno}, J.; {Ruiz Cobo}, B.
\newblock {An open-source, massively parallel code for non-LTE synthesis and
  inversion of spectral lines and Zeeman-induced Stokes profiles}.
\newblock {\em Astron. Astrophys.} {\bf 2015}, {\em 577},~A7,
\newblock
  doi:10.1051/0004-6361/201424860.

\bibitem[{Elsworth} \em{et~al.}(1994){Elsworth}, {Howe}, {Isaak}, {McLeod},
  {Miller}, {New}, {Speake}, and {Wheeler}]{elsworth94}
{Elsworth}, Y.; {Howe}, R.; {Isaak}, G.R.; {McLeod}, C.P.; {Miller}, B.A.;
  {New}, R.; {Speake}, C.C.; {Wheeler}, S.J.
\newblock {Solar Seismology---The Velocity Continuum Spectrum}.
\newblock {\em Mon. Not. R. Astron. Soc.} {\bf 1994}, {\em 269},~529,
\newblock
  doi:10.1093/mnras/269.3.529.

\bibitem[{Pall{\'e}} \em{et~al.}(1999){Pall{\'e}}, {Roca Cort{\'e}s},
  {Jim{\'e}nez}, {GOLF Team}, and {Virgo Team}]{palle99}
{Pall{\'e}}, P.L.; {Roca Cort{\'e}s}, T.; {Jim{\'e}nez}, A.; {GOLF Team};
  {Virgo Team}.
\newblock {The Sun as a Star: Background, Intensity and Velocity, Power Spectra
  and Convection}.
\newblock  In \emph{Stellar Structure: Theory and Test of Connective Energy Transport}; {  Astronomical Society of the Pacific Conference Series};
  {Gimenez}, A., {Guinan}, E.F., {Montesinos}, B., Eds.; {\bf 1999}; {Astronomical Society of the Pacific; San Francisco, CA, USA} Volume 173, p. 297.
  
\bibitem[{Queloz} \em{et~al.}(2001){Queloz}, {Henry}, {Sivan}, {Baliunas},
  {Beuzit}, {Donahue}, {Mayor}, {Naef}, {Perrier}, and {Udry}]{queloz01}
{Queloz}, D.; {Henry}, G.W.; {Sivan}, J.P.; {Baliunas}, S.L.; {Beuzit}, J.L.;
  {Donahue}, R.A.; {Mayor}, M.; {Naef},~D.; {Perrier},~C.; {Udry}, S.
\newblock {No planet for HD 166435}.
\newblock {\em Astron. Astrophys.} {\bf 2001}, {\em 379},~279--287,
\newblock
  doi:10.1051/0004-6361:20011308.

\bibitem[{Povich} \em{et~al.}(2001){Povich}, {Giampapa}, {Valenti}, {Tilleman},
  {Barden}, {Deming}, {Livingston}, and {Pilachowski}]{povich01}
{Povich}, M.S.; {Giampapa}, M.S.; {Valenti}, J.A.; {Tilleman}, T.; {Barden},
  S.; {Deming}, D.; {Livingston}, W.C.; {Pilachowski}, C.
\newblock {Limits on Line Bisector Variability for Stars with Extrasolar
  Planets}.
\newblock {\em Astron. J.} {\bf 2001}, {\em 121},~1136--1146,
\newblock
  doi:10.1086/318745.

\bibitem[{Dall} \em{et~al.}(2006){Dall}, {Santos}, {Arentoft}, {Bedding}, and
  {Kjeldsen}]{dall06}
{Dall}, T.H.; {Santos}, N.C.; {Arentoft}, T.; {Bedding}, T.R.; {Kjeldsen}, H.
\newblock {Bisectors of the cross-correlation function applied to stellar
  spectra. Discriminating stellar activity, oscillations and planets}.
\newblock {\em Astron. Astrophys.} {\bf 2006}, {\em 454},~341--348,
\newblock
  doi:10.1051/0004-6361:20065021.

\bibitem[{Figueira} \em{et~al.}(2015){Figueira}, {Santos}, {Pepe}, {Lovis}, and
  {Nardetto}]{figueira15}
{Figueira}, P.; {Santos}, N.C.; {Pepe}, F.; {Lovis}, C.; {Nardetto}, N.
\newblock {Line-profile variations in radial-velocity measurements
  (Corrigendum). Two alternative indicators for planetary searches}.
\newblock {\em Astron. Astrophys.} {\bf 2015}, {\em 582},~C2,
\newblock
  doi:10.1051/0004-6361/201220779e.

\bibitem[{Lanza} \em{et~al.}(2018){Lanza}, {Malavolta}, {Benatti}, {Desidera},
  {Bignamini}, {Bonomo}, {Esposito}, {Figueira}, {Gratton}, {Scandariato},
  {Damasso}, {Sozzetti}, {Biazzo}, {Claudi}, {Cosentino}, {Covino}, {Maggio},
  {Masiero}, {Micela}, {Molinari}, {Pagano}, {Piotto}, {Poretti}, {Smareglia},
  {Affer}, {Boccato}, {Borsa}, {Boschin}, {Giacobbe}, {Knapic}, {Leto},
  {Maldonado}, {Mancini}, {Martinez Fiorenzano}, {Messina}, {Nascimbeni},
  {Pedani}, and {Rainer}]{lanza18}
{Lanza}, A.F.; {Malavolta}, L.; {Benatti}, S.; {Desidera}, S.; {Bignamini}, A.;
  {Bonomo}, A.S.; {Esposito}, M.; {Figueira},~P.; {Gratton}, R.; {Scandariato},
  G.; et al.
\newblock {The GAPS Programme with HARPS-N at TNG. XVII. Line profile
  indicators and kernel regression as diagnostics of radial-velocity variations
  due to stellar activity in solar-like stars}.
\newblock {\em Astron. Astrophys.} {\bf 2018}, {\em 616},~A155,
\newblock
  doi:10.1051/0004-6361/201731010.

\bibitem[{Sulis} \em{et~al.}(2017{\natexlab{a}}){Sulis}, {Mary}, and
  {Bigot}]{sulis17}
{Sulis}, S.; {Mary}, D.; {Bigot}, L.
\newblock {A Study of Periodograms Standardized Using Training Datasets and
  Application to Exoplanet Detection}.
\newblock {\em IEEE Trans. Signal Proc.} {\bf 2017}, {\em
  65},~2136--2150, 
\newblock
  doi:10.1109/TSP.2017.2652391.

\bibitem[{Sulis} \em{et~al.}(2017{\natexlab{b}}){Sulis}, {Mary}, and
  {Bigot}]{sulis17b}
{Sulis}, S.; {Mary}, D.; {Bigot}, L.
\newblock {A Bootstrap Method for Sinusoid Detection in Colored Noise and
  Uneven Sampling. Application to Exoplanet Detection}.
\newblock In Proceedings of the {2017 25th European Signal Processing Conference (EUSIPCO)}, 28 August--2 September; Kos, Greece {\bf 2017}; pp. 1095--1099.

\bibitem[{Beckers}(2007)]{beckers07}
{Beckers}, J.M.
\newblock {Can variable meridional flows lead to false exoplanet detections?}
\newblock {\em Astron. Nachr.} {\bf 2007}, {\em 328},~1084,
\newblock
  doi:10.1002/asna.200710830.

\bibitem[{Makarov}(2010)]{makarov10}
{Makarov}, V.V.
\newblock {Variability of Surface Flows on the Sun and the Implications for
  Exoplanet Detection}.
\newblock {\em Astrophys.~J.} {\bf 2010}, {\em 715},~500--505,
\newblock
  doi:10.1088/0004-637X/715/1/500.

\bibitem[{Cegla} \em{et~al.}(2012){Cegla}, {Watson}, {Marsh}, {Shelyag},
  {Moulds}, {Littlefair}, {Mathioudakis}, {Pollacco}, and {Bonfils}]{cegla12}
{Cegla}, H.M.; {Watson}, C.A.; {Marsh}, T.R.; {Shelyag}, S.; {Moulds}, V.;
  {Littlefair}, S.; {Mathioudakis}, M.; {Pollacco}, D.; {Bonfils}, X.
\newblock {Stellar jitter from variable gravitational redshift: Implications
  for radial velocity confirmation of habitable exoplanets}.
\newblock {\em Mon. Not. R. Astron. Soc.} {\bf 2012}, {\em 421},~L54--L58,
\newblock
  doi:10.1111/j.1745-3933.2011.01205.x.

\end{thebibliography}

\reftitle{References}

\end{document}